\documentclass[aps,pra,showpacs,floatfix,superscriptaddress,11pt,nofootinbib]{revtex4-1}
\usepackage{xr}
\usepackage{amsmath,amssymb,physics}
\usepackage[toc,page]{appendix}
\usepackage{verbatim,graphicx,mathtools}
\usepackage{txfonts}
\usepackage{color}
\usepackage[all]{xy}
\usepackage{color}
\usepackage[all]{xy}
\usepackage{subfigure}
\usepackage{xcolor}
\usepackage{bm}
\usepackage{hyperref}

\newcommand{\nn}{\notag \\}
\newcommand{\be}{\begin{eqnarray}}
\newcommand{\ee}{\end{eqnarray}}

\newcommand{\bmat}{\left ( \begin{array}{cc} }
	\newcommand{\emat}{\end{array} \right ) }

\newcounter{jvct}

\newcommand{\beq}{\begin{equation}}
\newcommand{\beqs}{\begin{equation*}}
\newcommand{\eeq}{\end{equation}}
\newcommand{\eeqs}{\end{equation*}}

\allowdisplaybreaks

\begin{document}

\title{Phase diagram of a two-site coupled complex SYK model}
\author{Antonio M. Garc\'ia-Garc\'ia}
\email{amgg@sjtu.edu.cn}
\affiliation{Shanghai Center for Complex Physics, 
	School of Physics and Astronomy, Shanghai Jiao Tong
	University, Shanghai 200240, China}

\author{Jie Ping Zheng}
\email{jpzheng@sjtu.edu.cn}
\affiliation{Shanghai Center for Complex Physics, 
	School of Physics and Astronomy, Shanghai Jiao Tong
	University, Shanghai 200240, China}

\author{Vaios Ziogas}
\email{vaios.ziogas@sjtu.edu.cn}
\affiliation{Shanghai Center for Complex Physics, 
	School of Physics and Astronomy, Shanghai Jiao Tong
	University, Shanghai 200240, China}

\begin{abstract} 
	We study the thermodynamic properties of a two-site coupled complex Sachdev-Ye-Kitaev (SYK) model in the large $N$ limit by solving the saddle-point Schwinger-Dyson (SD) equations. We find that its phase diagram is richer than in the Majorana case. In the grand canonical ensemble, we identify a region of small chemical potential, and weak coupling between the two SYKs, for which two first order thermodynamic phase transitions occur as a function of temperature. First, we observe a transition from a cold wormhole phase to an intermediate phase that may correspond to a charged wormhole. For a higher temperature, there is another first order transition to the black hole phase. As in the Majorana case, the low temperature wormhole phase is gapped and, for sufficiently large coupling between the two complex SYK, or chemical potential, the first order transitions become crossovers. The total charge is good indicator to study the phase diagram of the model: it is zero in the cold wormhole phase and jumps discontinuously at the temperatures at which the transitions take place. Based on the approximate conformal symmetry of the ground state, expected to be close to a thermofield double state, we identify the effective low energy action of the model. It is a generalized Schwarzian action with $SL(2,R)\times U(1)$ symmetry with an additional potential and a extra degree of freedom related to the charge. In the large $N$ limit, results from this low energy action are consistent with those from the solution of the SD equations. Our findings are a preliminary step towards the characterization of traversable wormholes by its field theory dual, a strongly interacting fermionic system with charger, that is easier to model experimentally.  
\end{abstract}\maketitle

\newpage
\section{{Introduction}}
Traversable wormholes are classical solutions of Einstein's equations  representing shortcuts in the geometry that may allow tele-transportation among distant regions of space-time. For that reason, it has been a recurrent research theme for several decades \cite{morris1988,visser1989,hawking1988,visser1989a,hawking1990}.
Unfortunately, these solutions are considered classically unphysical as matter in these backgrounds would violate some fundamental physical principle, such as the null energy condition.

The situation changed recently \cite{gao2016}, after it was shown that
turning on an interaction that couples the two boundaries of an eternal BTZ black hole, the quantum matter stress tensor has a negative average null
energy without violating causality which suggests that the wormhole becomes traversable. Other examples in different backgrounds and dimensionalities, but with similar features, were found shortly after \cite{maldacena2017,marlof2018,Tumurtushaa:2018agq,maldacena2018b,Anabalon:2018rzq}. 
The next main development came after Maldacena and Qi \cite{maldacena2018}, (see also \cite{bak2018} that performed an explicit bulk time evolution), constructed a near
${\rm AdS}_2$ background whose ground state was a time independent traversable wormhole, termed {\it eternal traversable wormhole}.
Its ground state is highly entangled and close to a thermofield double
state (TFD) \cite{israel1976,maldacena2003} that it separated from the first excited state by a gap. For sufficiently weak coupling, the system has a first order transition from the traversable wormhole phase to the (two) black hole phase. At a certain critical coupling, the gap vanishes and the transition becomes a crossover. The boundary theory is given by a generalized Schwarzian action related to a modified Liouville quantum mechanical problem. Unlike the standard SYK case \cite{bagrets2016,bagrets2017}, the spectrum for low energies is discrete, representing the wormhole phase. However, for higher energies, it is continuous, representing the black hole phase.  

Interestingly, the field theory dual of this eternal traversable wormhole was identified \cite{maldacena2018} to be the low energy phase of a two-site coupled Sachdev-Ye-Kitaev model. It can be shown that the real time evolution of this model leads to the formation of a traversable wormhole \cite{maldacena2019}. Previously, a non-random SYK model \cite{ferrari2017} was conjectured to describe similar physics. Indeed, both models share the same pattern of symmetry breaking \cite{kim2019,klebanov2020}.

The SYK model \cite{bohigas1971,bohigas1971a,french1970,french1971,mon1975,benet2003,kota2014,sachdev1993,sachdev2010,kitaev2015,fu2018,jensen2016,jevicki2016} is a toy model for holography that, based on the same pattern of symmetry breaking from the full conformal group to SL(2,R) \cite{maldacena2016a,almheiri2015}, is believed to be dual of a certain near ${\rm AdS}_2$ background \cite{maldacena2016a}.  Its main interest is that, 
despite being strongly interacting, and quantum chaotic \cite{kitaev2015}, it is analytically tractable \cite{kitaev2015}. Distinctive features of the model include: the saturation of a universal bound on chaos \cite{maldacena2015} typical of fast scramblers of information and systems with a gravity dual \cite{maldacena2015}, the exponential growth \cite{maldacena2016,cotler2016,garcia2017} of low energy excitations, typical of quantum black holes, and spectral correlations described by random matrix theory \cite{garcia2016,cotler2016,garcia2017,Altland:2017eao,garcia2018a} that suggests that quantum dynamics is ergodic for sufficiently long times.
The high temperature limit of the two-site SYK model \cite{maldacena2018}, dual to two black hole backgrounds, shares most of these features. However, important differences arise in the low temperature limit corresponding to the wormhole phase. The system is no longer quantum chaotic. The low energy excitations are discrete even in the thermodynamics limit. This is consistent with the observation of a transition in level statistics \cite{garcia2019}, from integrable in the wormhole phase to quantum chaotic in the black hole region.

An appealing feature of the SYK model is that, at least potentially, it could be modeled experimentally \cite{danshita2016,chew2018,Pikulin2017}. That would allow not only the study of novel transport regimes in strongly interacting quantum dots but also, through holographic dualities,  the experimental test of certain aspects of quantum gravity. However, this program is hampered by the difficulty to isolate and handle Majorana fermions. 

In this paper, we study a generalization  of the two-site SYK model with Majorana fermions, dual of the eternal traversable wormhole \cite{maldacena2018}, to complex fermions with an extra $U(1)$ symmetry. More specifically, we shall study its expected field theory dual: a two-site coupled SYK model with Dirac, instead of Majoranas, fermions.
Single complex SYK have already been extensively investigated \cite{sachdev2015,davison2017,gu2020,sorokhaibam2020,sachdev2019} in the literature. Qualitatively, they retain most of the interesting features of the Majorana SYK model while being closer to more realistic models of strongly interacting electrons. 
We shall see that, to some extent, this applies to the coupled charged SYK model. We shall find that the model is still gapped for low temperatures and weak coupling which is a signature of the wormhole phase. For small chemical potential and low temperature, it has no charge which is another feature of the traversable wormhole dual to the Majorana two-site SYK model \cite{maldacena2018}. Likewise, the high temperature phase is consistent with that of a system whose gravity dual is two black holes.

However, as we increase the chemical potential, we have identified qualitative differences between the Majorana and complex cases in the grand canonical ensemble. In a relatively narrow range of parameters, there exists an intermediate phase between the cold wormhole and black hole phase. For weak coupling, this intermediate phase is separated from the black hole and cold wormhole phases by two first order transitions. Tentatively, we believe that this novel phase may be a charged wormhole  \cite{maldacena2019} characterized by a finite charge and a still gap in the spectrum. However, further research is required to confirm this point. For sufficiently large coupling, the transitions end in a crossover. 

The paper is organized as follows: in section \ref{sec:model} we introduce the model, its expected ground state and symmetries, and derive the Schwinger-Dyson equations for Green's functions in the large $N$ limit. Based on the numerical solution of these equations, section \ref{sec:fivenum} is devoted to a detailed analysis of the thermodynamic properties of the model and the resulting phase diagram. Based on the approximate conformal symmetry of the ground state and its soft breaking in the low temperature limit, in section \ref{sec:lowenergy}, we write down the low energy generalized Schwarzian effective action and study some of its properties. This is the region of parameters where a gravity dual may exist.
A list of problems for future research and conclusions are found in section \ref{sec:conclusion}. \\

\section{Coupled complex SYK model}\label{sec:model}
\subsection{Action}

The Hamiltonian of the complex SYK model is \cite{sachdev2015,davison2017}
\begin{align}\label{eq:HcomplexSYK}
H = \sum_{\{i\}} J_{i_1\dots i_q} \psi^{i_1\dagger}\dots\psi^{i_{q/2}\dagger} \psi^{i_{q/2+1}}\dots \psi^{i_q}\,,
\end{align}
where $\{i\}=\{1\leq i_1<i_2<\dots<i_{q/2}\leq N,\,1\leq i_{q/2+1}<\dots<i_{q}\leq N\}$ and the complex coupling $J$ satisfies
\begin{align}
J_{i_1\dots i_{q/2} i_{q/2+1}\dots i_q} & = J^{*}_{i_{q/2+1}\dots i_q i_1\dots i_{q/2}}\,, \quad  J_{i_1 i_2 \dots i_\frac{q}{2};i_{\frac{q}{2}+1}\dots i_q} = J_{[i_1 i_2 \dots i_\frac{q}{2}]; [i_{\frac{q}{2}+1}\dots i_q]}\,, \quad \overline{|J_{i_1\dots i_q}|^2} = \frac{\left(q/2\right)!^2}{2 q N^{q-1}} J^2\,,
\end{align}
where the overbar denotes a statistical average with zero mean.

The Hamiltonian $H$ has a global $U(1)$ conserved charge
\begin{align}\label{eq:globalu1_def}
Q = \frac{1}{N} \sum_i\psi^{i\dagger} \psi^i -\frac{1}{2}\,,
\end{align}
taking values in $(-1/2,1/2)$. We deform $H$ by adding a term $-N \mu Q$, where $\mu$ is the chemical potential of the $U(1)$ charge.

In this paper, we consider two such SYKs with Dirac fermions, which we call ``left'' $L$ and ``right'' $R$. The two Hamiltonians $H_L$ and $H_R$ involve the same realization of the disordered coupling $J_{i_1\dots i_q}$, and we set the chemical potentials to be equal, $\mu_L=\mu_R=\mu$, since we want to consider two identical copies of the theory. We couple the two systems using the interaction
\begin{align}\label{eq:kcoupl}
H_{int} =-\frac{1}{2 N^{p-1}} \left[ \sum_i\left(\eta\, \psi^{i\dagger}_{L}\psi^i_{R} + \eta^{*}\psi^{i\dagger}_{R}\psi^i_{L}\right)\right]^p\,,
\end{align}
where $\eta$ is in principle a complex coupling and $|\eta|\equiv \kappa$.\footnote{We can actually define $\eta$ with any phase if we perform a phase rotation in one of the systems, as explained in \cite{sahoo2020}, but we keep it general in our analytical treatment.} We choose $p=1$ for now, but it might also be interesting to consider $p>1$.
Therefore, the total Hamiltonian is given by, 
\begin{align}\label{hami}
H_{total}=H_L+H_R+H_{int}\,,
\end{align}
where $H_R=(-1)^{q/2} H_L$.

\subsection{Symmetries}

We first discuss the symmetries of the uncoupled model $\eta = 0$, 
\begin{align}\label{eq:uncoupled_def}
H_L +H_R -N\mu\, (Q_L+Q_R)\,.
\end{align}

There are two global symmetries $U(1)_L$ and $U(1)_R$, which can be combined as
\begin{align}\label{eq:u1+u1-}
U(1)_{\pm}=U(1)_L\pm U(1)_R\,, 
\end{align}
where $U(1)_{+}$ is interpreted as the symmetry of the system related to the total charge $Q_+ = Q_L + Q_R$. Note that, due to the form of \eqref{eq:uncoupled_def}, $\mu$ can be thought of as a chemical potential for $Q_+$.

If we totally antisymmetrize the fermions in both left and right site, we also have a particle-hole symmetry at zero chemical potential $\mu=0$  \cite{Fu:2016yrv,Gu:2019jub}:
\begin{align}\label{eq:particle-hole_symm}
\psi_{a}^{i} \leftrightarrow \psi^{i\dagger}_{a}\,, \qquad J_{i_1\dots i_{q/2} i_{q/2+1}\dots i_q} \rightarrow J^{*}_{i_1\dots i_{q/2} i_{q/2+1}\dots i_q}\,,
\end{align}
for $a=L,R$. The extra terms coming from the antisymmetrization are subleading in the $N$ expansion, so they do not affect our large-$N$ analytic arguments.

There is also an interchange symmetry
\begin{align}\label{eq:interchange_symm}
\psi_{L}^{i} \leftrightarrow \psi^{i}_{R}\,.
\end{align}

A general combination of $U(1)_L\times U(1)_R$ and the interchange symmetry \eqref{eq:interchange_symm} gives
\begin{align}\label{eq:u1u1int}
\psi_{L}^{i} \rightarrow e^{i\theta_R}\psi^{i}_{R}\,,\qquad \psi_{R}^{i} \rightarrow e^{i\theta_L}\psi^{i}_{L}\,.
\end{align}

The 2-fermion coupling \eqref{eq:kcoupl} breaks $U(1)_{-}$ explicitly down to $\mathbb{Z}_2$, while it preserves $U(1)_{+}$.\footnote{In \cite{klebanov2020,sahoo2020}, the $U(1)_{-}$ was broken spontaneously by a quartic coupling, and \cite{klebanov2020} interpreted it as corresponding to a bulk gauge field in the gravity dual theory.} Demanding that the combination \eqref{eq:u1u1int} is preserved leads to the following relation between the fermionic phase in each site,
\begin{align}\label{eq:int_coupl_sym}
\theta_R-\theta_L = 2 \theta\,, \qquad \theta\equiv\arg\eta\,.
\end{align}
We make the convenient choice $\theta_{R}=-\theta_{L}=\theta$.\footnote{Alternatively, \eqref{eq:int_coupl_sym} leads to the $\mathbb{Z}_4$ symmetry $\psi_{L}^{i} \rightarrow \psi^{i}_{R}\,,\, \psi_{R}^{i} \rightarrow -\psi^{i}_{L}$ being preserved upon choosing $\theta_R=0\,,\,\theta_L=\pi$ and $\eta$ purely imaginary as in \cite{sahoo2020}.}

\subsection{Schwinger-Dyson equations}
One of our main goals is to study the thermodynamic properties of the system. For that purpose, the first step is to derive the saddle point equations, termed Schwinger-Dyson (SD) equations \cite{maldacena2016}, that control the large $N$ limit of Green's functions and self-energies that enter in the calculation of thermodynamic quantities.    
We first perform the statistical average of the path integral for the total action related to the Hamiltonian (\ref{hami}). To leading order in $N$, a straightforward calculation \cite{bagrets2016} yields,  

\begin{align}\label{eq:path_int2}
\langle Z\rangle_J &= \int D\psi^{i\dagger}_a D\psi^i_a \exp\Bigl[ -\sum_{i,a}\int\psi^{i\dagger}_{a}\left(\partial_{\tau} - \mu\right) \psi^i_{a} +\sum_i\int\left(\eta\,\psi^{i\dagger}_{L}\psi^i_{R} + \eta^{*}\psi^{i\dagger}_{R}\psi^i_{L}\right) \nn
&+\frac{\left(-1\right)^{q/2}}{q N^{q-1}} J^2 \sum_{a,b} \int\int s_{ab}\, \left(|\sum_{i}\psi^{i\dagger}_a(\tau) \psi^{i}_b(\tau')|^2\right)^{q/2} \Bigr]\,,
\end{align}
where $a=L,R$ and $s_{LL}=s_{RR}=1,\,s_{LR}=s_{RL}=(-1)^{q/2}$.

We now define the Green's functions,
\begin{align}\label{eq:greens_def}
G_{ab}(\tau,\tau')=\frac{1}{N}\sum_{i} \langle\mathcal{T}\psi^{i\dagger}_a(\tau) \psi^{i}_b(\tau')\rangle\,,
\end{align}
obeying
\begin{align}\label{eq:herm_greens}
G_{ab}^{*}(\tau,\tau') = G_{ba}(\tau,\tau')\,.
\end{align}
The interchange symmetry \eqref{eq:u1u1int}, \eqref{eq:int_coupl_sym} implies that
\begin{align}\label{eq:int_sym_greens}
G_{LR}(\tau,\tau') = e^{-2i\theta} G_{RL}(\tau,\tau')\,,
\end{align}
or $G_{LR} = - G_{RL}$ when $\theta=\pi/2$.

	The particle-hole symmetry \eqref{eq:particle-hole_symm} implies that
	\begin{align}\label{eq:p-h_greens}
	G_{ab}(\tau,\tau') = - G_{ba}(\tau,\tau')\,,
	\end{align}
	at zero chemical potential $\mu=0$. Combining \eqref{eq:p-h_greens} with \eqref{eq:herm_greens} gives 
	\begin{align}\label{eq:greens_ph_relations}
	\textrm{Re}G_{LL}=\textrm{Re}G_{RR}\,,\qquad \textrm{Im}G_{LR}=-\textrm{Im}G_{RL}\,,\qquad \textrm{Im}G_{LL}=\textrm{Im}G_{RR}=\textrm{Re}G_{LR}=\textrm{Re}G_{RL}=0\,.
	\end{align}
	This is for reference only, we do not assume these symmetries in the following as we want to work at $\mu\neq0$.

At finite temperature $T=1/\beta$, the KMS condition (for the thermofield double state) reads \cite{doi:10.1142/3277}
\begin{align}\label{eq:KMS}
G_{ab}(\tau) = - G_{ab}(\tau + \beta)\,,\qquad \tau<0\,.
\end{align}

Introducing in the path integral \eqref{eq:path_int2} the Lagrange multipliers $\Sigma_{ab}(\tau,\tau')$ which enforce the definition \eqref{eq:greens_def}, and integrating out the fermions assuming a replica-diagonal ansatz, we find
\begin{align}
\langle Z\rangle_J &\sim \int D G\, D\Sigma\, \exp \left[ -N\,I_{eff}\right]\,,
\end{align}
where

\begin{align}\label{eq:effective_action_cSYK}
-I_{eff} &= \log \det \left[ \delta_{ab}\left(\partial_{\tau} - \mu\right) +\Sigma_{ab}\right] + \int\left[\eta\,G_{LR}(\tau,\tau) + \eta^{*}G_{RL}(\tau,\tau)\right] \nn
&+\sum_{a,b} \int\int \left[\Sigma_{ba}(\tau',\tau)G_{ab}(\tau,\tau') +\left(-1\right)^{q/2} q^{-1}J^2 s_{ab}\, \left[G_{ab}(\tau,\tau')\right]^{q/2} \left[G_{ba}(\tau',\tau)\right]^{q/2} \right] \,.
\end{align}

Varying the effective action \eqref{eq:effective_action_cSYK} with respect to $G_{ab}$ and $\Sigma_{ab}$ leads to the following saddle-point SD equations:
\begin{align}\label{eq:c_SDequations}
&\tilde{\Sigma}_{LL}(\tau,\tau') = -\left(-1\right)^{q/2} J^2 \left[G_{LL}(\tau,\tau')\right]^{q/2} \left[G_{LL}(\tau',\tau)\right]^{q/2-1} +\left(\partial_{\tau} - \mu\right)\delta(\tau-\tau')\,,\nn
&\tilde{\Sigma}_{LR}(\tau,\tau') = -\left(-1\right)^{q} J^2 \left[G_{LR}(\tau,\tau')\right]^{q/2} \left[G_{RL}(\tau',\tau)\right]^{q/2-1}{+\eta}\, \delta(\tau-\tau')\,,\nn
&\tilde{\Sigma}_{LL}\star G_{LL} (\tau,\tau') +\tilde{\Sigma}_{LR}\star G_{RL} (\tau,\tau')= -\delta(\tau-\tau') \,,\nn
&\tilde{\Sigma}_{LL}\star G_{LR} (\tau,\tau') +\tilde{\Sigma}_{LR}\star G_{RR} (\tau,\tau')= 0 \,,
\end{align}
where the star $\star$ denotes the convolution $ \left(f\star g\right) (\tau_1,\tau_2) \equiv \int d\tau f(\tau_1,\tau) g(\tau,\tau_2)$ and we have defined
\begin{align}
\tilde{\Sigma}_{ab}(\tau,\tau')=\Sigma_{ab}(\tau,\tau') +\left(\partial_{\tau} - \mu\right){\delta}(\tau-\tau')\delta_{ab}\,.
\end{align}

We also get another set of equations by exchanging $L\leftrightarrow R$ and $\eta\leftrightarrow\eta^{*}$ in \eqref{eq:c_SDequations}. For convenience, in the numerical calculation of thermodynamic properties, we set $\eta=-i\kappa$ and $q = 4$.

These equations \eqref{eq:c_SDequations} are the analogous  
to those for the two-site coupled Majorana SYK model
\cite{maldacena2018,maldacena2019} whose gravity dual in the limit of low temperature and weak-coupling between left and right sites is the eternal traversable wormhole.

\subsection{The grand potential}


In the large N limit, the grand potential, computed by inserting the solution of the SD equations \eqref{eq:c_SDequations} in the on-shell action \eqref{eq:effective_action_cSYK}, is given by
\begin{align}\label{eq:grand_potential}
-\frac{\beta\Omega}{N}&=2\ln [2\cosh\frac{\beta\mu}{2}] + Tr\ln\frac{(i\omega+\mu-\Sigma_{LL})(i\omega+\mu-\Sigma_{RR})-\Sigma_{LR}\Sigma_{RL}}{(i\omega+\mu)(i\omega+\mu)}\nn
& -(-1)^{q/2}(1-\frac{1}{q}) \sum_{ab}\int d\tau \left(J^2G_{ab}(\tau)^{q/2}G_{ba}(\beta-\tau)^{q/2}\right)\,,
\end{align}
where we have regularized the determinant as in \cite{Gu:2019jub}. From this expression, we can compute the rest of thermodynamic quantities.

\section{Thermodynamic properties in the large N limit and phase diagram}\label{sec:fivenum}
In this section, we investigate the thermodynamic properties of the Hamiltonian (\ref{hami}) in the grand canonical ensemble which will result in a detailed phase diagram of the model as a function of the coupling $\kappa$ between the two complex SYKs and the chemical potential $\mu$. We obtain all the thermodynamic quantities of interest by solving numerically the SD equations (\ref{eq:c_SDequations}) using standard iterative techniques.

We will be mostly interested in the calculation of the total charge $Q$, related to the global $U(1)$ symmetry mentioned earlier and the grand potential $\Omega(T,\kappa,\mu)$ (\ref{eq:grand_potential}) where $\kappa = |\eta|$. We shall see that these quantities characterize the different phases of the model and are easily accessible from the knowledge of the Green's functions and self-energies resulting from the solution of the SD equations (\ref{eq:c_SDequations}). Likewise, the analysis of the exponential decay of $G_{LR}$ will provide useful information on the gap $E_g$ that characterizes the wormhole phase. We initiate our analysis with the calculation of the grand potential $\Omega$ (\ref{eq:grand_potential}).

\begin{figure}
	\subfigure[]{\includegraphics[scale=.33]{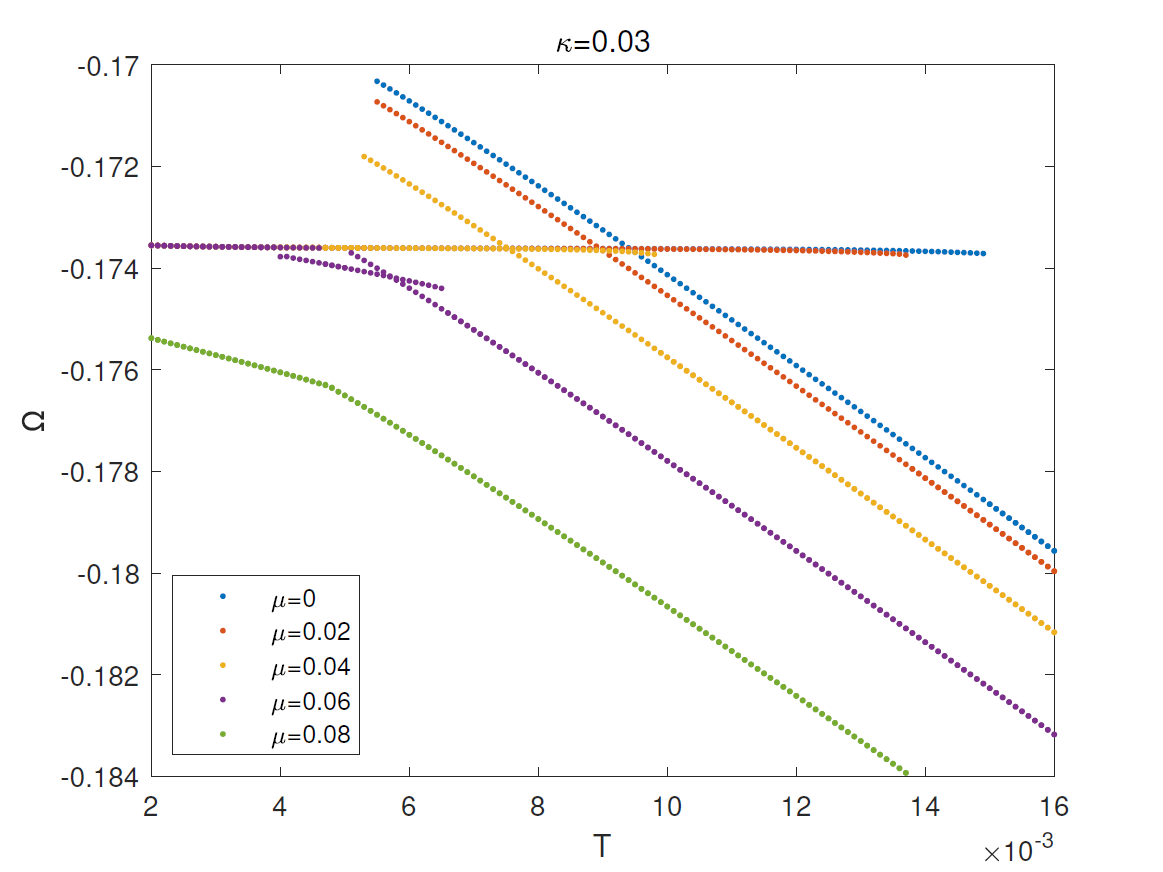}}
	\subfigure[]{\includegraphics[scale=.33]{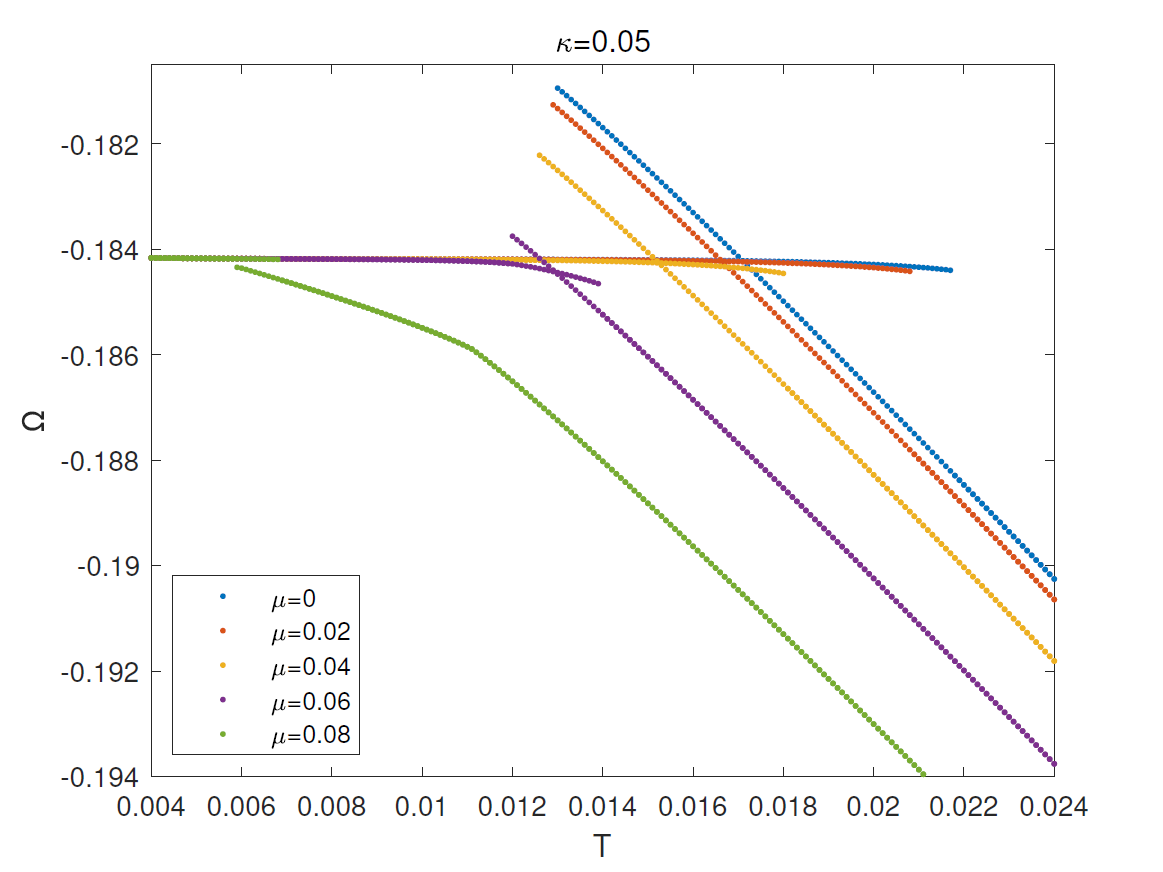}}
	\subfigure[]{\includegraphics[scale=.33]{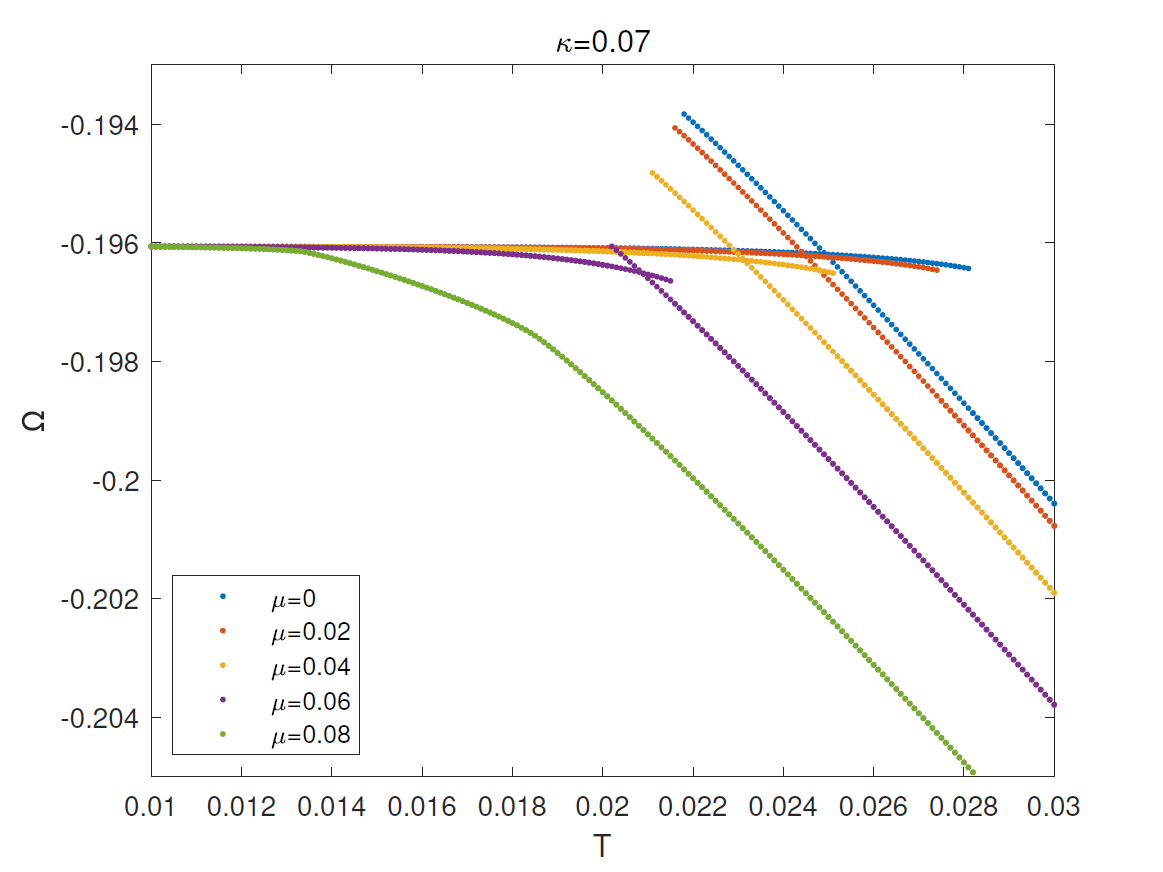}}
	\subfigure[]{\includegraphics[scale=.33]{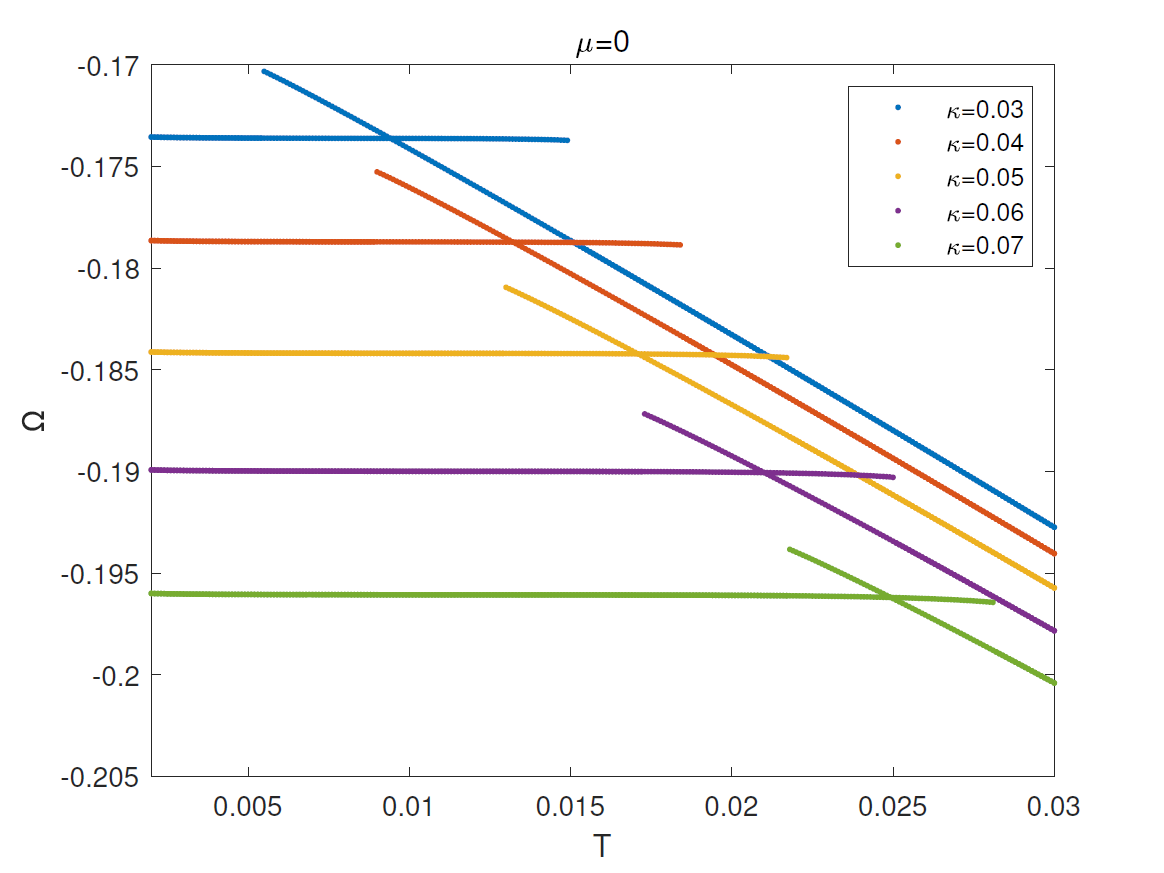}}
	\subfigure[]{\includegraphics[scale=.33]{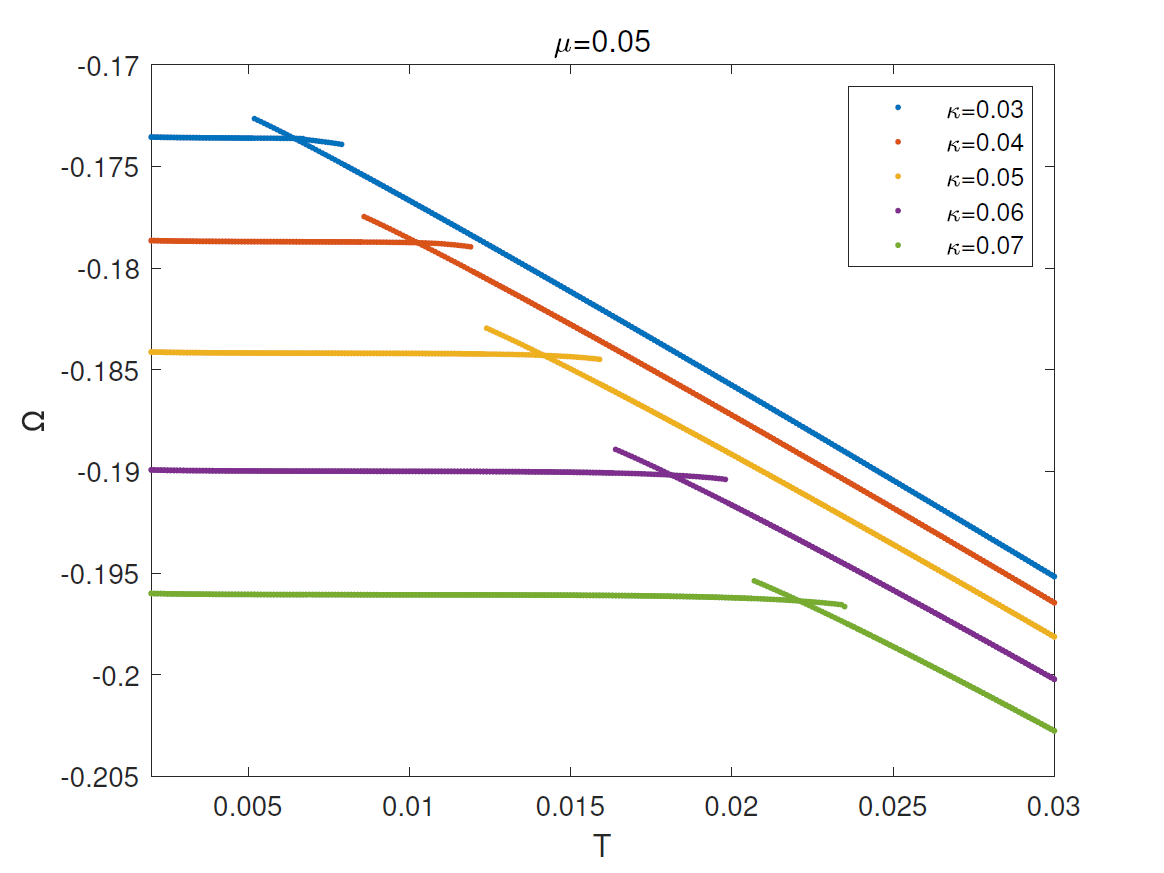}}
	\subfigure[]{\includegraphics[scale=.33]{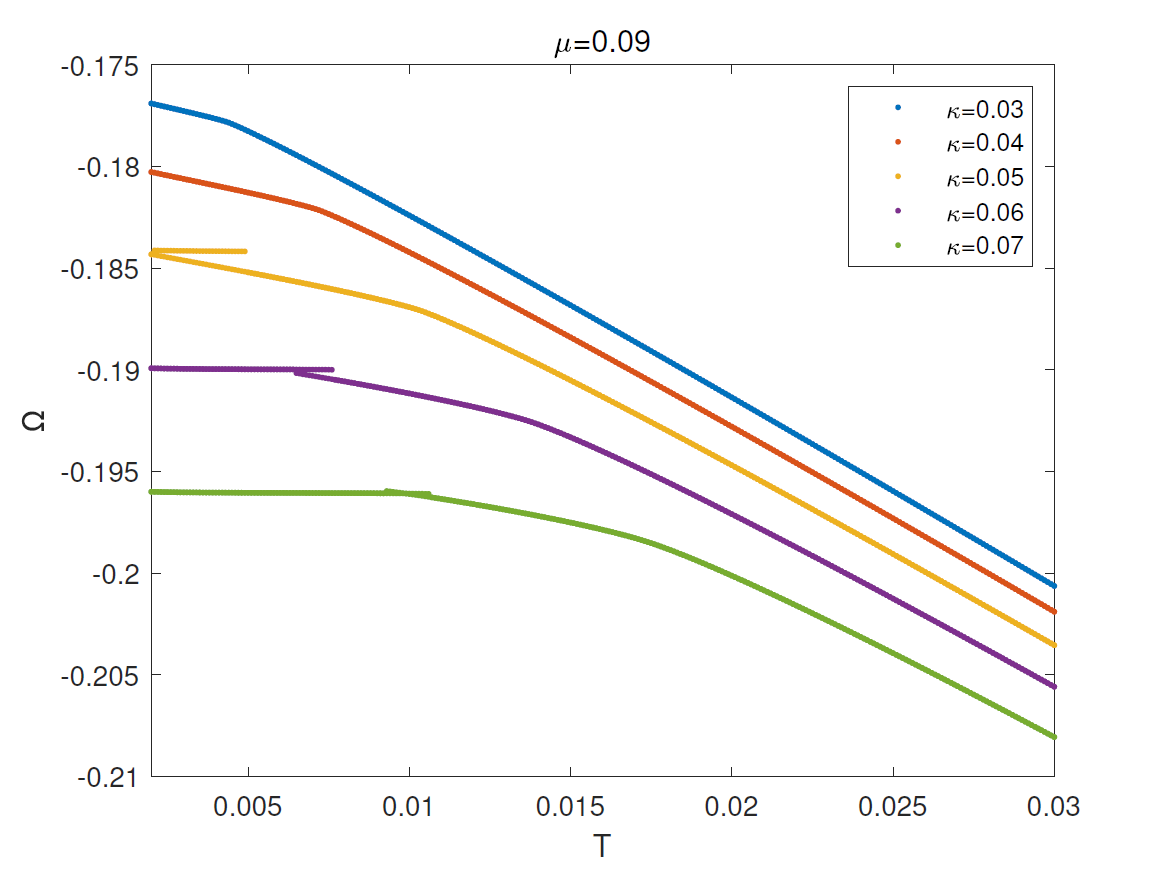}}
	\caption{Grand potental $\Omega$ versus temperature $T$ for $\kappa=0.03$ (a), $\kappa=0.05$ (b), $\kappa=0.07$ (c), with $\mu=0,~0.02,~0.04,~0.06,~0.08$ and for $\mu=0$ (d), $\mu=0.05$ (e), $\mu=0.09$ (f), with $\kappa=0.03,~0.04,~0.05,~0.06,~0.07$. The almost constant grand potential is a signature of the wormhole phase. A finite $\mu$ suppresses the wormhole phase and eventually induces a new phase transition for sufficiently low temperatures. As was expected, the increase of $\kappa$, for a fixed $\mu$, enhances the wormhole phase. }\label{fig:Omg_kp_mu}
\end{figure}

\begin{figure}
	\includegraphics[scale=.5]{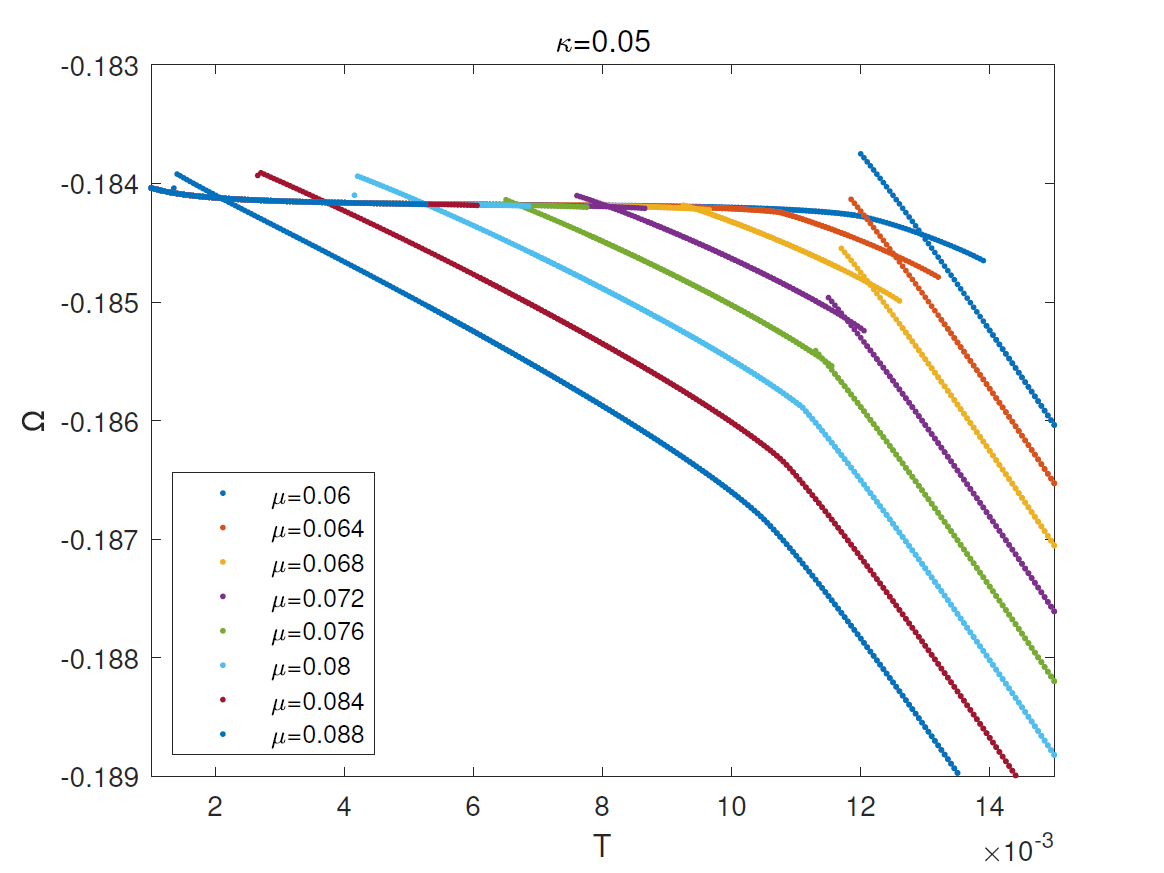}
	\caption{Grand potental $\Omega$ versus temperature $T$ in the region of parameters  $\kappa=0.05$ with $\mu=0.06, \cdots, 0.088$ where the two first order phase transitions are more clearly observed. We observe that the transitions become crossovers for sufficiently large $\mu$. }\label{fig:Omg_kp_05H}
\end{figure}

\subsection{Grand potential $\Omega(T,\kappa,\mu)$}

We compute the grand potential $\Omega$ as a function of the temperature $T$ for various $\kappa$ and $\mu$ by plugging in the action the Green's function and self-energies obtained by the numerical solution of the SD equations (\ref{eq:c_SDequations}). The final expression for the grand potential, after a determinant regularization \cite{gu2020}, is given by (\ref{eq:grand_potential}). 

In general, the SD equations for a given temperature can have more than one solution corresponding to different phases of the model. The preferred solution is the one with a lower value of the grand potential. 

Figure \ref{fig:Omg_kp_mu} depicts the temperature dependence of the grand potential for several couplings $\kappa$ (upper plots) and chemical potentials $\mu$ (lower plots). For $\mu = 0$, results are very similar to the Majorana case \cite{maldacena2018}. For very low temperatures, and finite but small $\kappa$, the grand potential is almost temperature independent suggesting the existence of a gap in the spectrum. This is the expected behavior in the traversable wormhole phase where the ground state, approximately described by a zero entropy TFD state, is separated from the first excited by a energy gap. As temperature increases, we observe a kink indicating a first order transition. In the proximity of the transition, we show the two branches of the grand potential. The gravity dual of the higher temperature phase is expected to be a two black hole geometry with an explicit coupling between the two backgrounds. For sufficiently strong coupling $\kappa \geq 0.125$, the first order transition ends. It is replaced by a smooth crossover of no evident gravitational interpretation. 

The situation becomes more interesting for finite but small $\mu$. For sufficiently small $\mu$, the grand potential is similar to the $\mu=0$ case. We still observe a flat low temperature grand potential, related to a gap in the spectrum typical of the wormhole phase, that eventually ends in the first order transition mentioned above. The only difference is that  $\mu$ slightly lowers both the gap and the critical temperature. This is expected as physically, the chemical potential effectively increases the energy of the system that is detrimental of the gap in the wormhole phase which will vanish at a lower temperature. 

However, for $\mu \sim 0.05$ we start to observe a second transition around $\kappa = 0.05$. More specifically, as temperature increases, the wormhole phase undergoes a first order transition to an intermediate phase. At a higher temperature, another first order transition occurs from this intermediate phase to the black hole phase. That the transition to the two black hole phase is the one at higher temperature can be inferred from the slope of the grand potential which is very similar for all values of the chemical potential no matter whether the intermediate phase exists or not. When $\mu$ increases further, this novel transition becomes also a crossover so we can identify regions with two transitions, two crossovers and one transition and one crossover. In general, the window of parameters where the two transitions are observed is rather narrow.

In
figure \ref{fig:Omg_kp_05H}, we choose the optimal choice of parameters for which this second transition is more clearly observed. We note that despite the second transition occurring in a relatively small range of parameters, the region of coexistence of the different phases, given by the range of temperatures in which other branches are present, is still much smaller than the range of temperatures in which the intermediate phase occurs.  This is a strong indication that this phase is stable in the grand canonical ensemble.

A technical comments is in order. As was mentioned earlier, the different solutions of the SD equations only exist in a determined range of temperatures. As the end points of each branch are approximated, the numerical calculation becomes increasing unstable with larger convergence time. We cannot rule out that a given branch survives for a somehow larger range of temperatures. 

\begin{figure}
	\subfigure[]{\includegraphics[scale=.36]{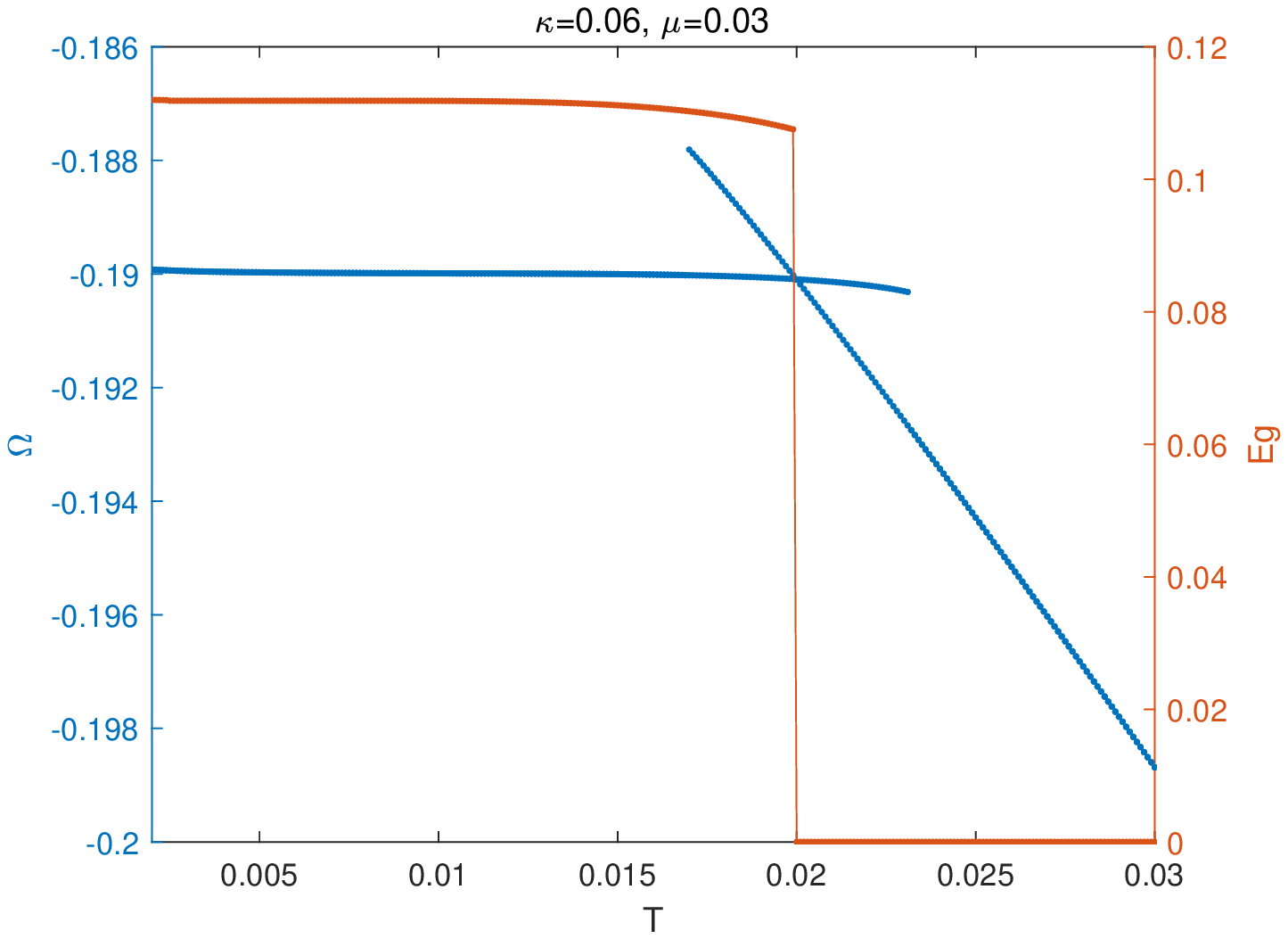}}
	\subfigure[]{\includegraphics[scale=.36]{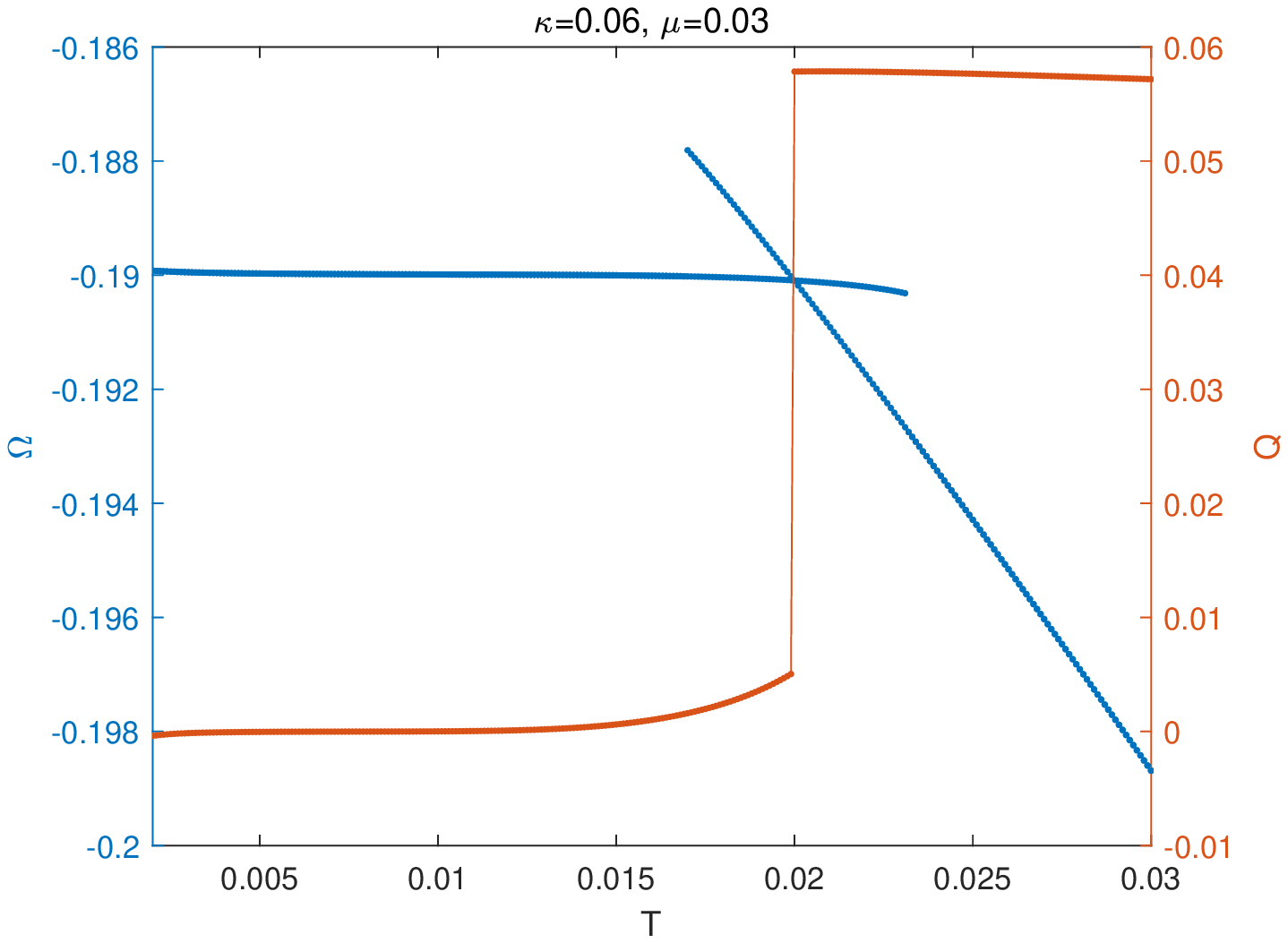}}
	\subfigure[]{\includegraphics[scale=.36]{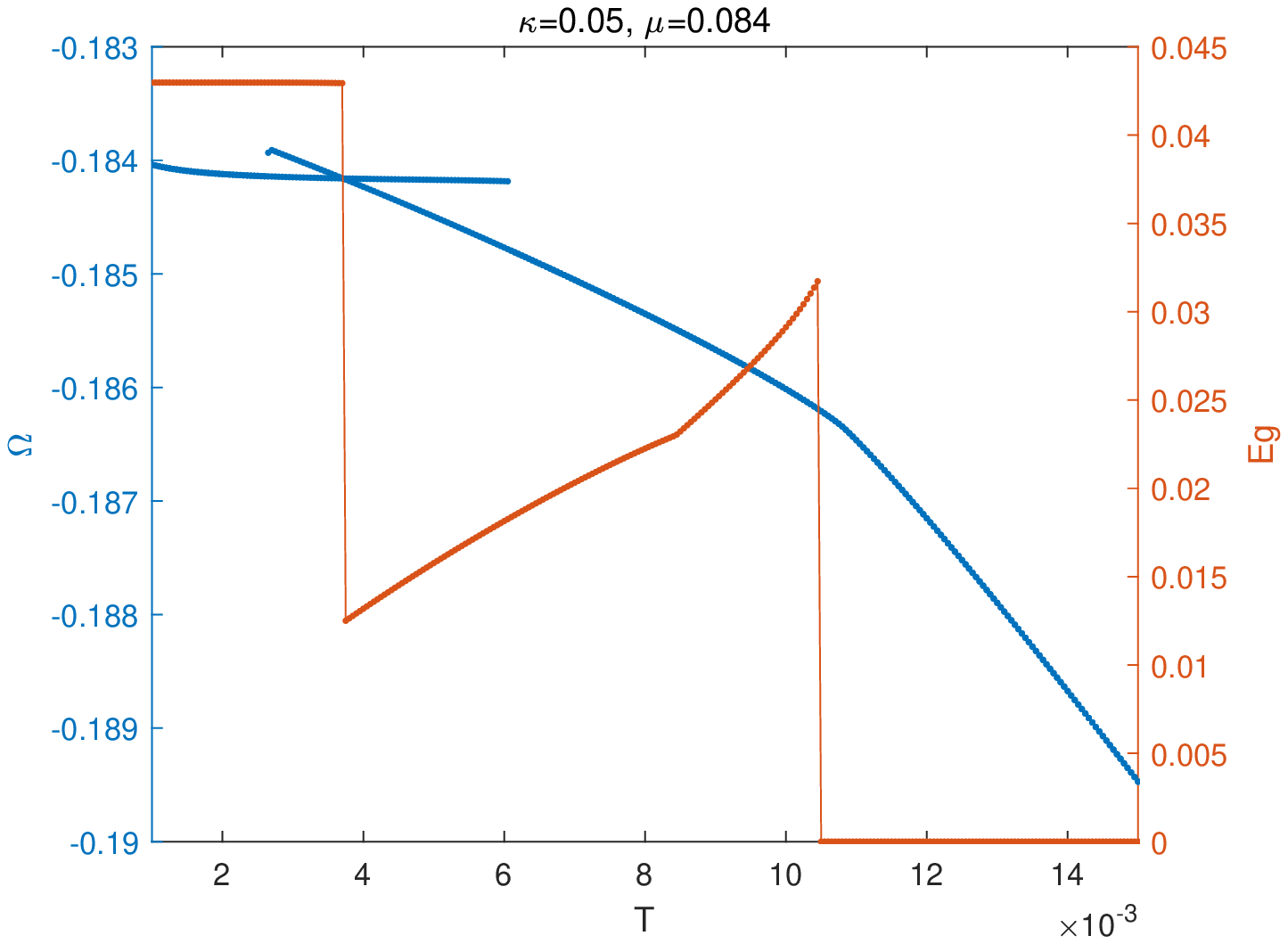}}
	\subfigure[]{\includegraphics[scale=.36]{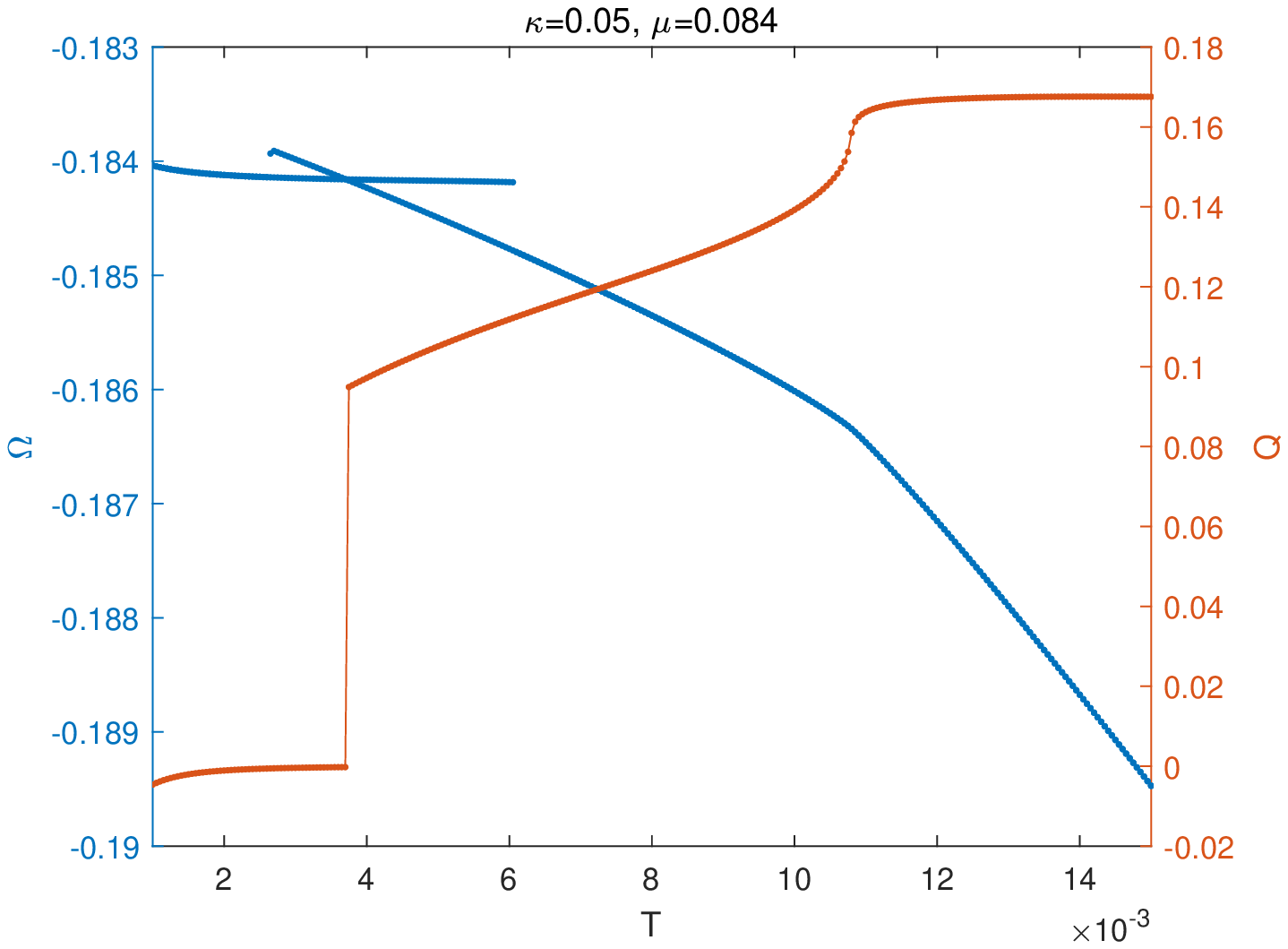}}
	\subfigure[]{\includegraphics[scale=.36]{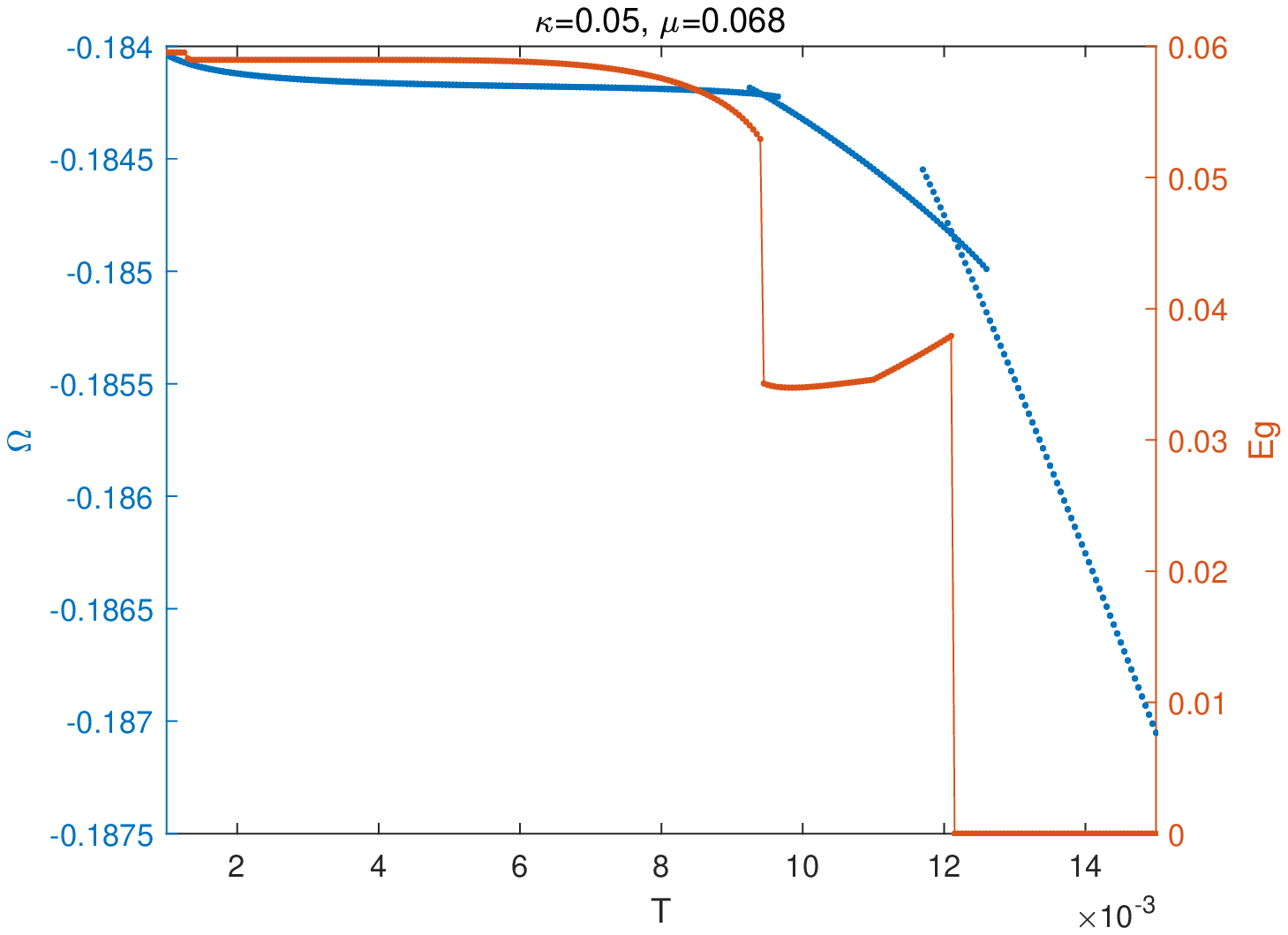}}
	\subfigure[]{\includegraphics[scale=.36]{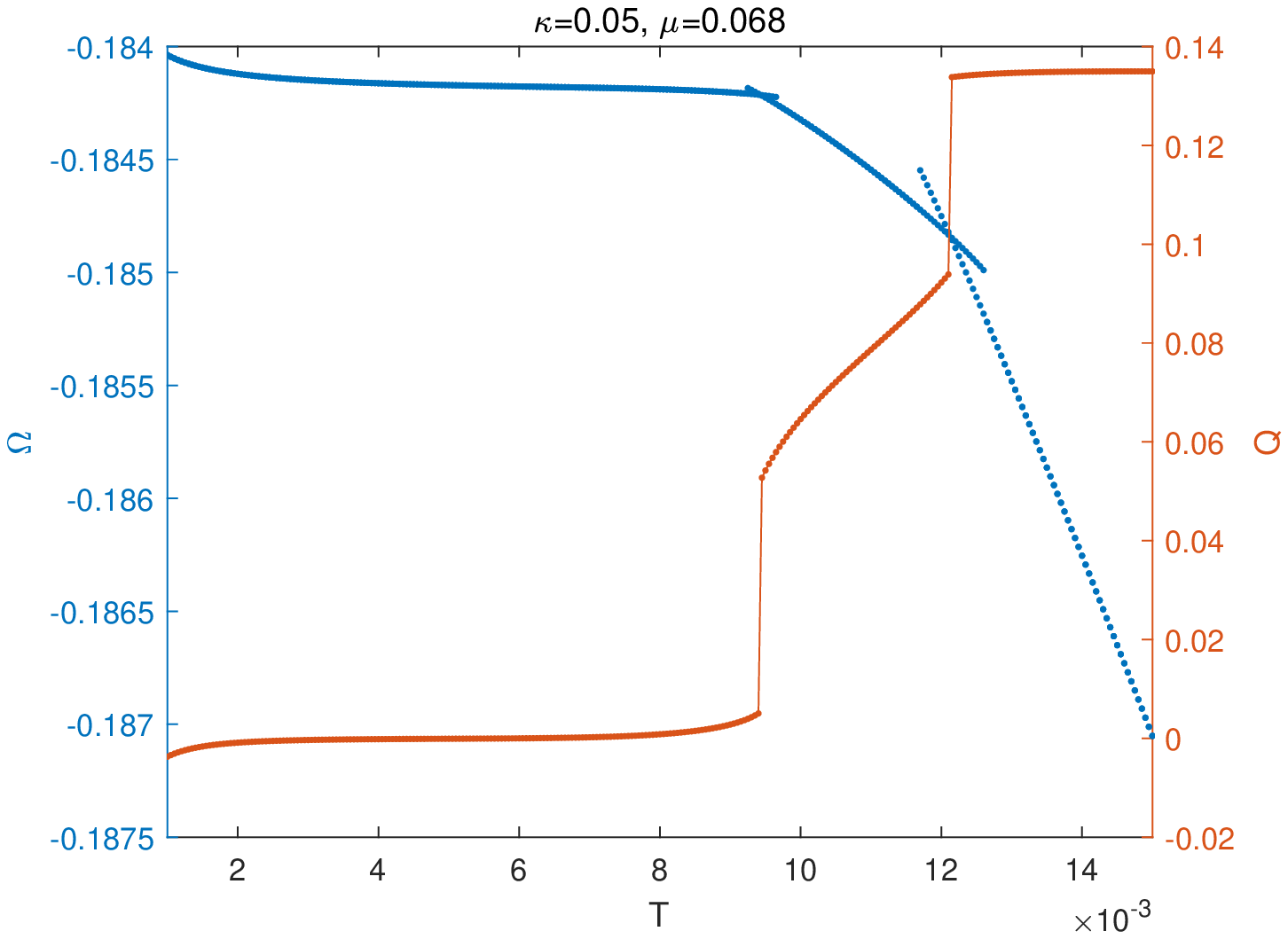}}
	\subfigure[]{\includegraphics[scale=.36]{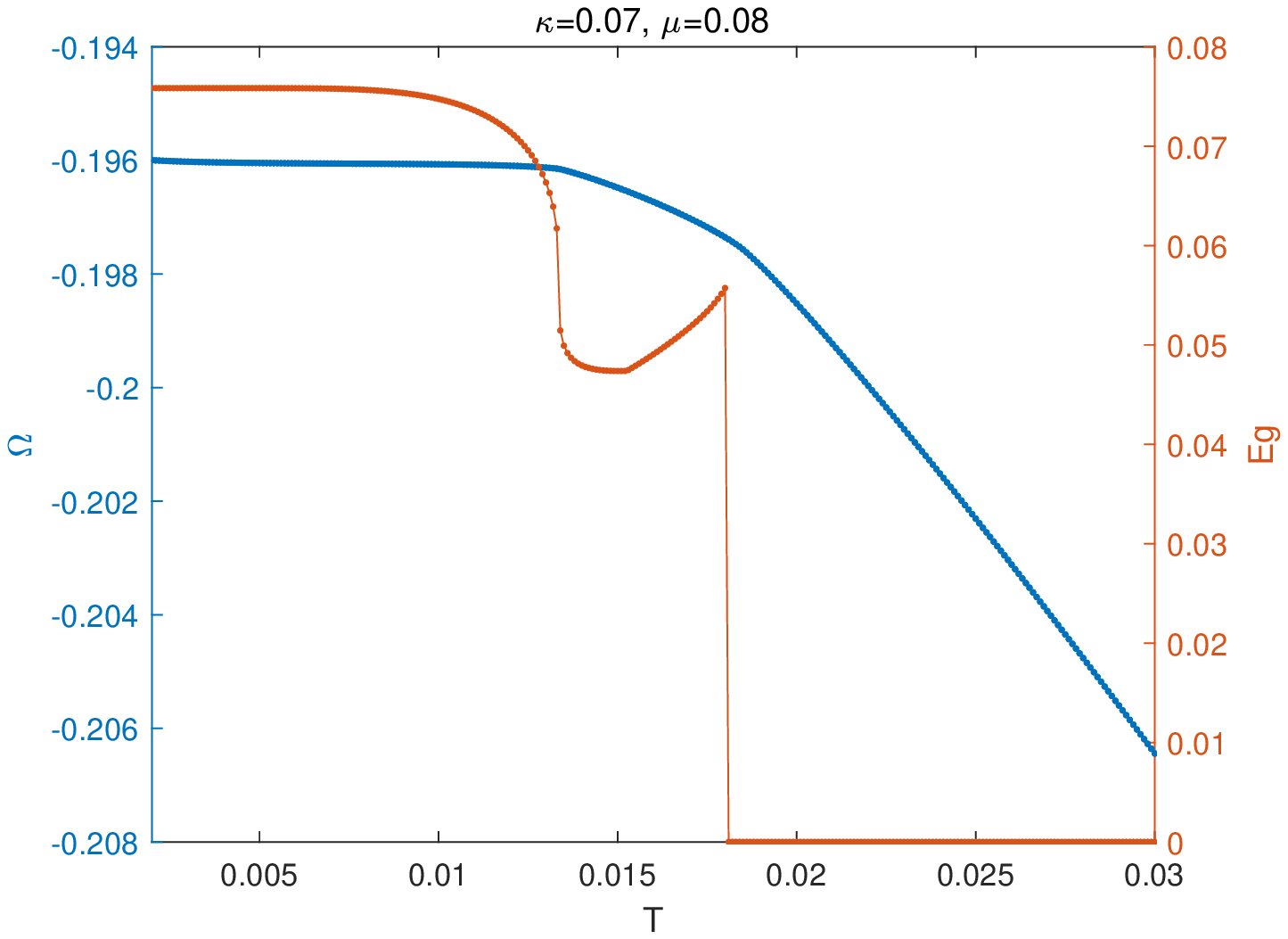}}
	\subfigure[]{\includegraphics[scale=.36]{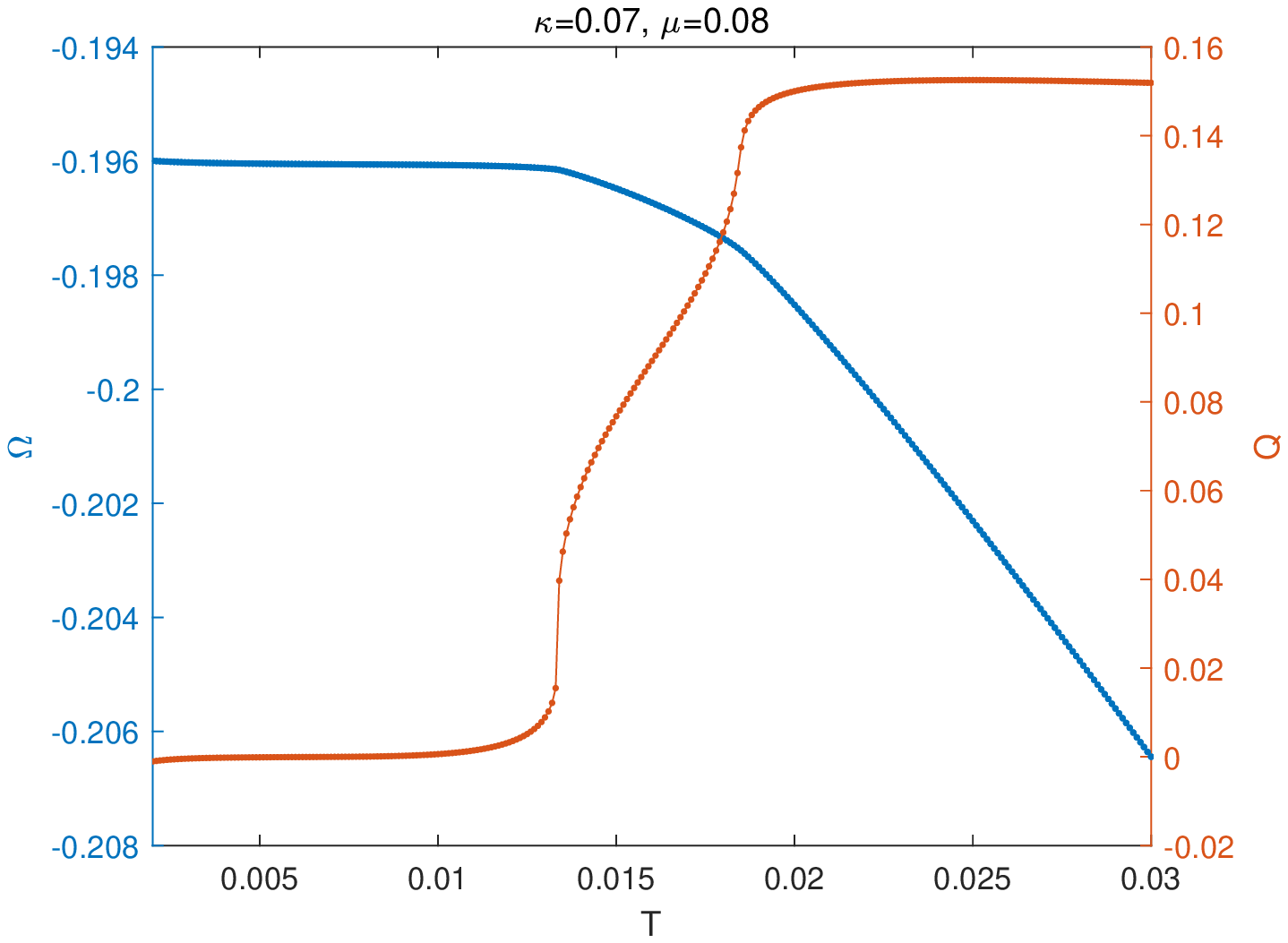}}
	\caption{$\Omega(T)$ and $E_g(T) $ (left column) and $\Omega(T)$ and $Q(T)$ (right column) for different $\mu, \kappa$. (a),(b): Only one phase transition, wormhole to black hole, is observed. (c),(d): only the lower temperature transition is observed, the high temperature one becomes a crossover. The fit to obtain the gap deteriorates rapidly for temperatures above the low temperature transition so results are less reliable; (e),(f): two phase transitions are observed. The fit to obtain the gap in the intermediate region is less reliable; (g),(h): for sufficiently large $\mu, \kappa$ both transitions becomes crossovers.}\label{fig:transition}
\end{figure}
\begin{figure}
	\centering
	\includegraphics[scale=.49]{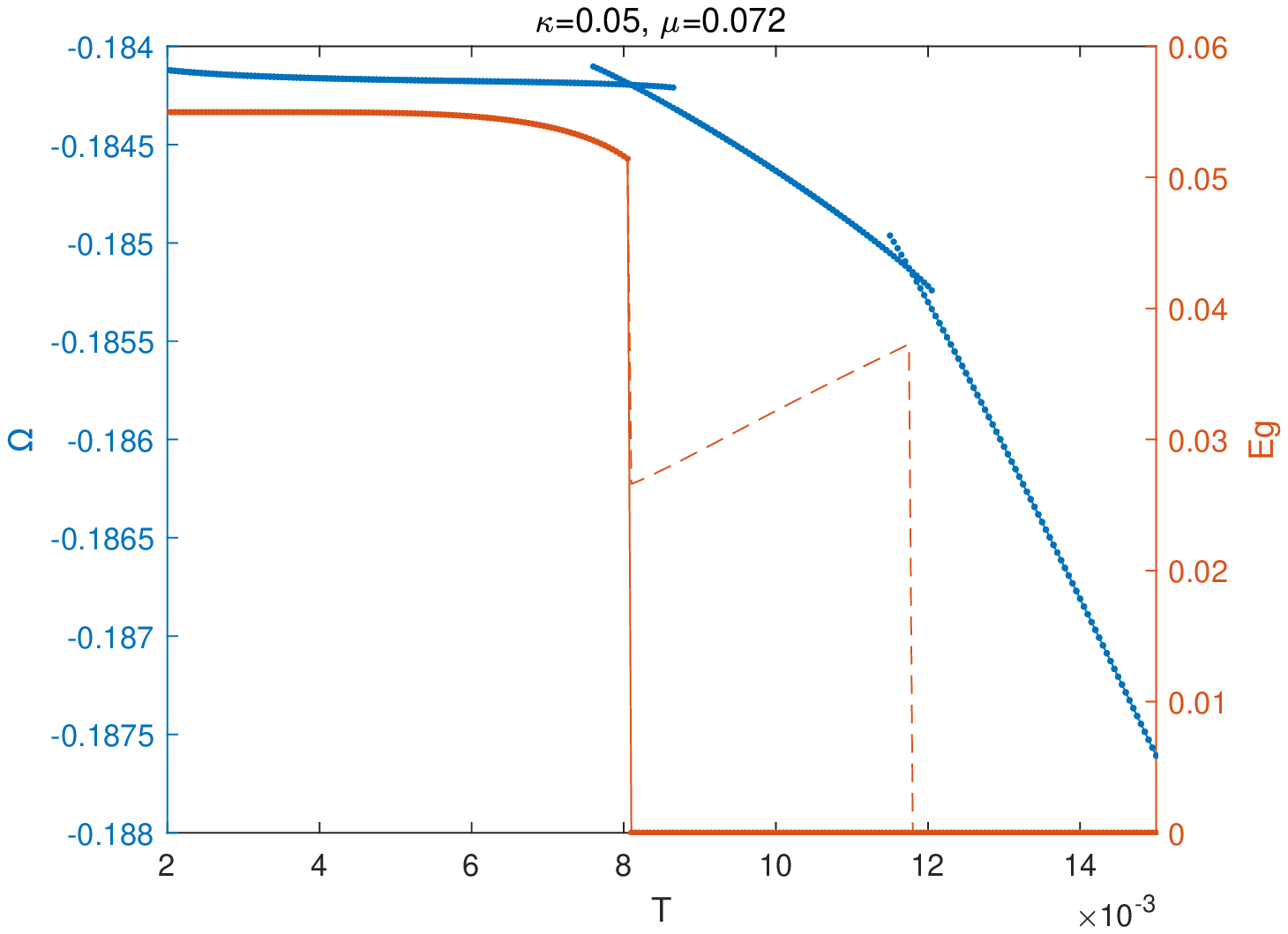}
	\includegraphics[scale=.49]{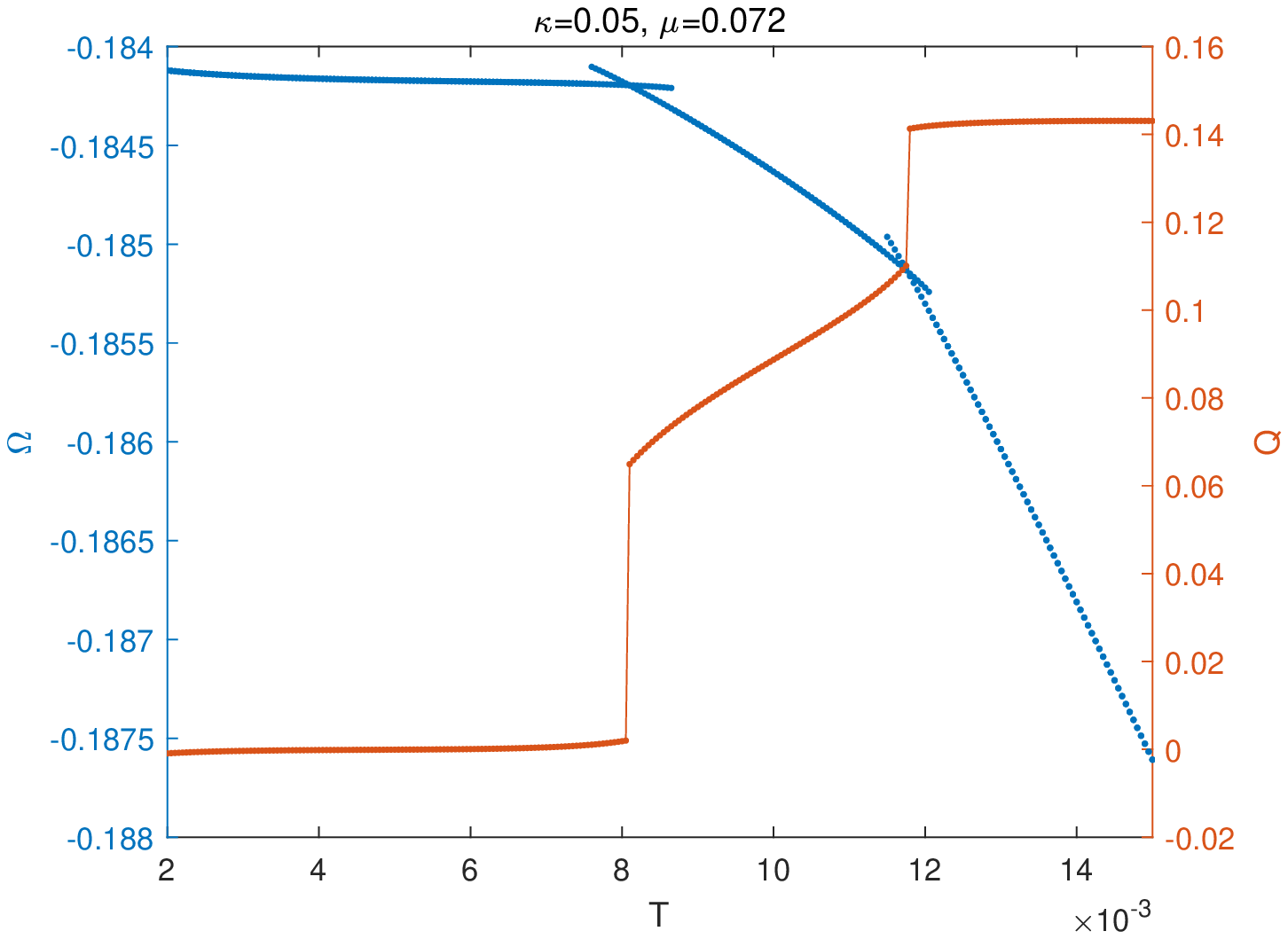}
	\caption{Left: $\Omega(T)$ and $E_g(T) $ for $\kappa=0.05$ and $\mu=0.072$ where the two transitions are clearly observed. The solid line is the gap obtained by a linear fit of $\ln G_{LR}$. The dashed line is a similar fit but the exponential decay of $G_{LR}$ occurs in a shorter interval so the fitting is less reliable. Right: $\Omega(T)$ and $Q(T)$. In the wormhole low temperature phase,  charge is zero. It jumps at both the low temperature and high temperature transitions. It increases linearly between the two transitions. This linear increase, related to a constant entropy, may be a feature of the  intermediate charged wormhole phase.}\label{fig:Eg_Q}
\end{figure}

\subsection{The energy gap $E_g$ and the charge $Q$}
In order to further elucidate the phase diagram of the model, especially the nature of the intermediate phase, we study the energy gap and the charge $Q$ related to the global $U(1)$ symmetry mentioned in previous sections. We start with a detailed introduction of both concepts.
\subsubsection{The charge $Q$}
Following \eqref{eq:globalu1_def}, we define the charge, 
\begin{equation}
Q_{aa}=\frac{1}{N}\sum_i^N \psi_{ai}^\dagger \psi_{ai} - \frac{1}{2},
\end{equation}
where the index $a = L,R$.  
Recalling the definition of the Green function above, 
\begin{equation}
G_{ab}(\tau)=\frac{1}{N}\sum_{i}^N\langle T\psi_{ai}^\dagger(\tau)\psi_{bi}(0) \rangle
\Rightarrow\left\{\begin{aligned}
& G_{ab}(\epsilon)=\frac{1}{N}\sum_i^N\langle \psi_{ai}^\dagger(\epsilon)\psi_{bi}(0) \rangle  \\
& G_{ab}(-\epsilon)=-\frac{1}{N}\sum_i^N\langle \psi_{bi}(0)\psi_{ai}^\dagger(-\epsilon) \rangle
\end{aligned}\right.\end{equation}
we can express the charge as a function of this Green's function,  
\begin{equation}\begin{aligned}
Q_{aa}= & \frac{1}{N}\sum_i^N\langle \psi_{ai}^\dagger\psi_{ai}\rangle-\frac{1}{2} =\lim_{\epsilon\to 0} \frac{1}{N}\sum_i^N\langle \psi_{ai}^\dagger(\epsilon)\psi_{ai}(0)\rangle-\frac{1}{2} =G_{aa}(0^+)-\frac{1}{2}  \\
=& \frac{1}{2}-\frac{1}{N}\sum_i^N\langle \psi_{ai}\psi_{ai}^\dagger\rangle =\frac{1}{2}-\lim_{\epsilon\to 0} \frac{1}{N}\sum_i^N\langle \psi_{ai}(0)\psi_{ai}^\dagger(-\epsilon)\rangle =\frac{1}{2}+G_{aa}(0^-)=\frac{1}{2}-G_{aa}(\beta-0^+)
\end{aligned}\end{equation}
leading to \cite{gu2020}, 
\begin{equation}
Q_{aa}=\frac{1}{2}(G_{aa}(0^+)-G_{aa}(\beta-0^+)).
\end{equation}

Since $G_{LL}(\tau)=G_{RR}(\tau)$ is real, we define the total charge $Q$ as\footnote{This charge was called $Q_{+}$ in the previous sections.}
\begin{equation}
Q=Q_{LL}+Q_{RR}= Q_+ = G_{LL}(0^+)-G_{LL}(\beta-0^+).
\end{equation}



The temperature dependence of the charge is illustrated in the right column of figure \ref{fig:transition} for a broad range of parameters. Interestingly, it trails with great accuracy the transition in the grand potential. It is almost temperature independent, and very close to zero, in the low temperature wormhole phase. Physically, it means that at sufficiently low temperatures the wormhole ground state is robust to the presence of a small chemical potential. Both, a finite temperature or chemical potential increase the energy of the system but the interactions are strong enough to balance these increases and keep the charge almost zero. Only the energy gap decreases as $\mu$ increases. 

At the two transitions, for small $\kappa$ and $\mu$, the charge jumps so it can be employed to detect the transition in the system. For larger values of $\kappa$ and $\mu$, the transition in the grand potential becomes a crossover. In this range of parameters, the abrupt discontinuous changes in the charge become sharp but smooth so the study of $Q$ provides a rather detailed knowledge of the phase diagram of the model.

In figure \ref{fig:Eg_Q} (right column), we show the temperature dependence of $Q$ for a choice of parameters where the two phase transitions are observed with special clarity. In the low temperature phase, the charge is close to zero, but at the low temperature phase transition it jumps to a finite value. In the intermediate phase, the charge increases approximately linearly with temperature so in the intermediate region the entropy is constant. At the high temperature phase transition, towards the black hole phase, it jumps again. It remains an open question if the jump in the charge in the low temperature phase transition, is associated with a qualitative change in the wormhole ground state, namely, it is unclear whether the wormhole geometry is robust and becomes charged or the finite charge signal a transition to a background with no traversability. 
The calculation of the energy gap will shed some light on this issue.   
\begin{figure}
	\subfigure[]{\includegraphics[scale=.33]{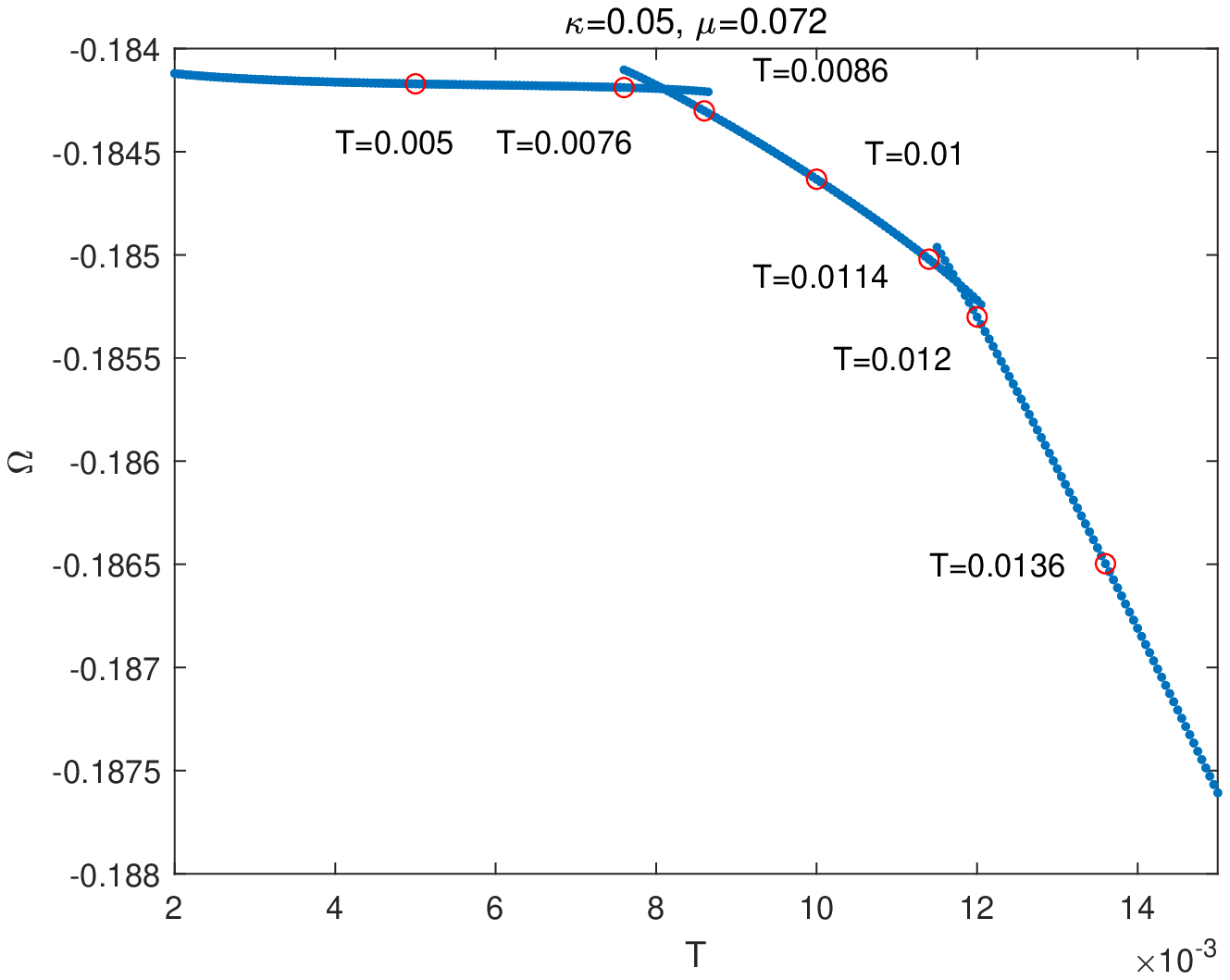}}
	\subfigure[]{\includegraphics[scale=.33]{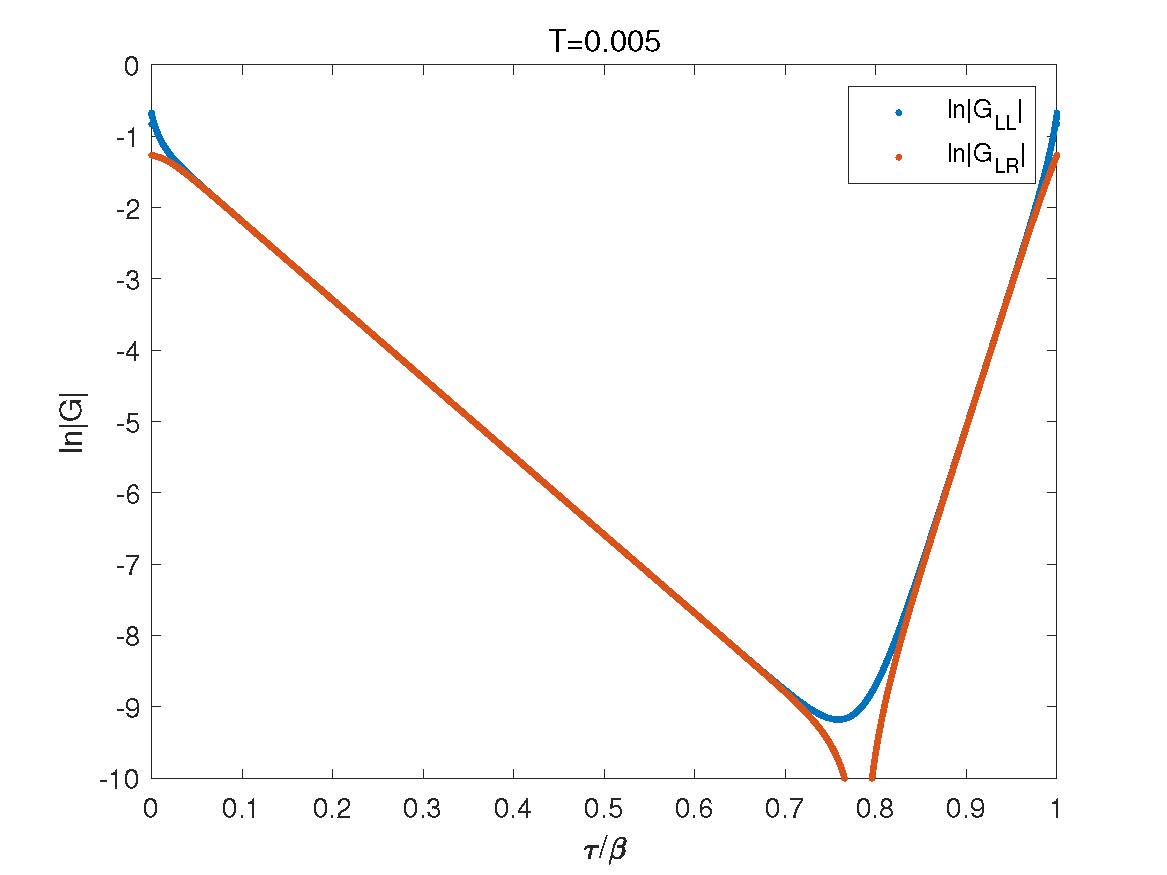}}
	\subfigure[]{\includegraphics[scale=.33]{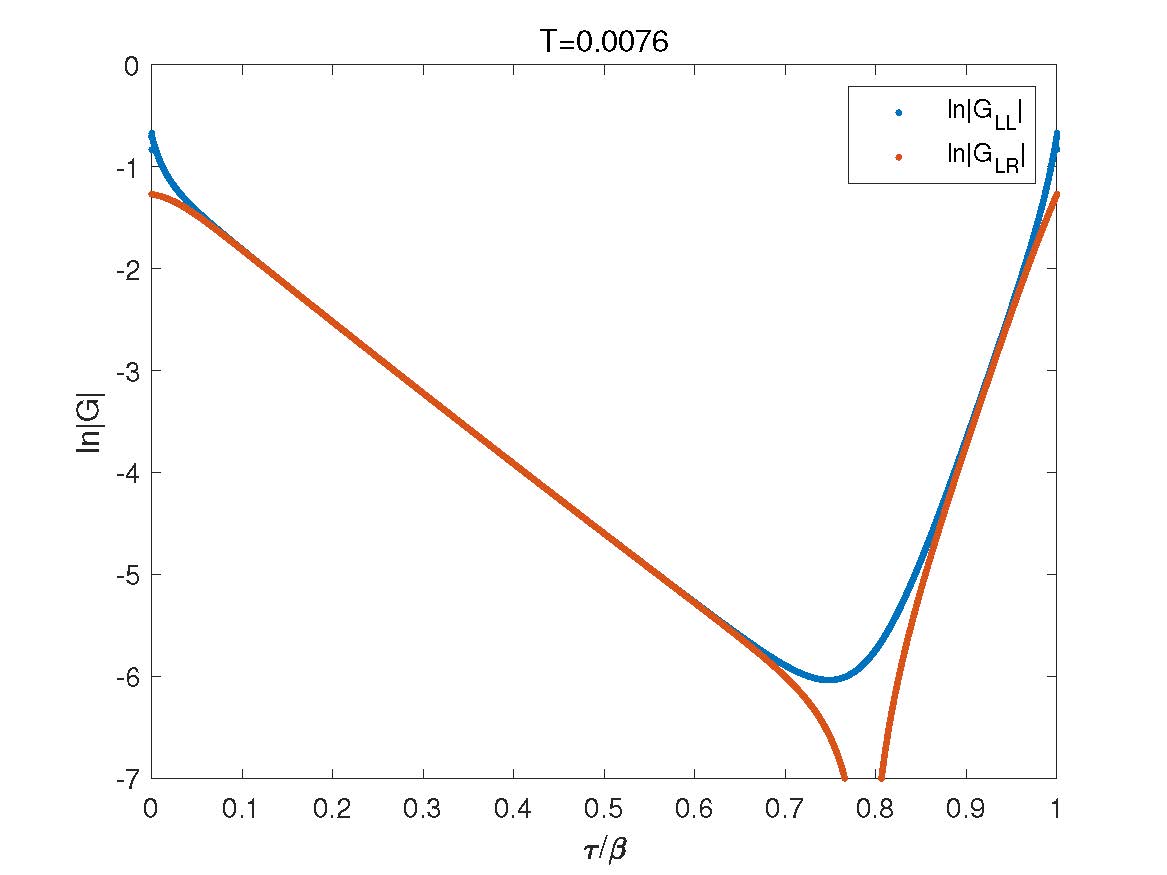}}
	\subfigure[]{\includegraphics[scale=.33]{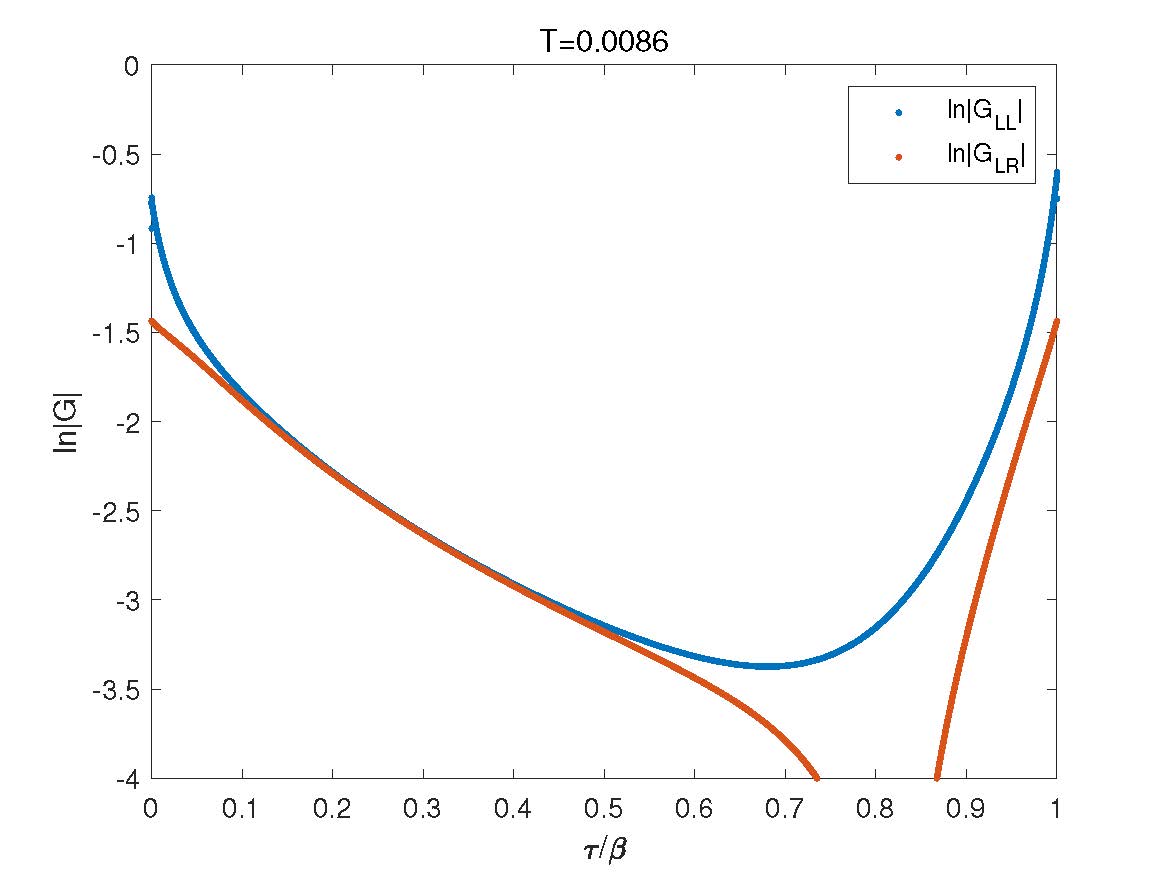}}
	\subfigure[]{\includegraphics[scale=.33]{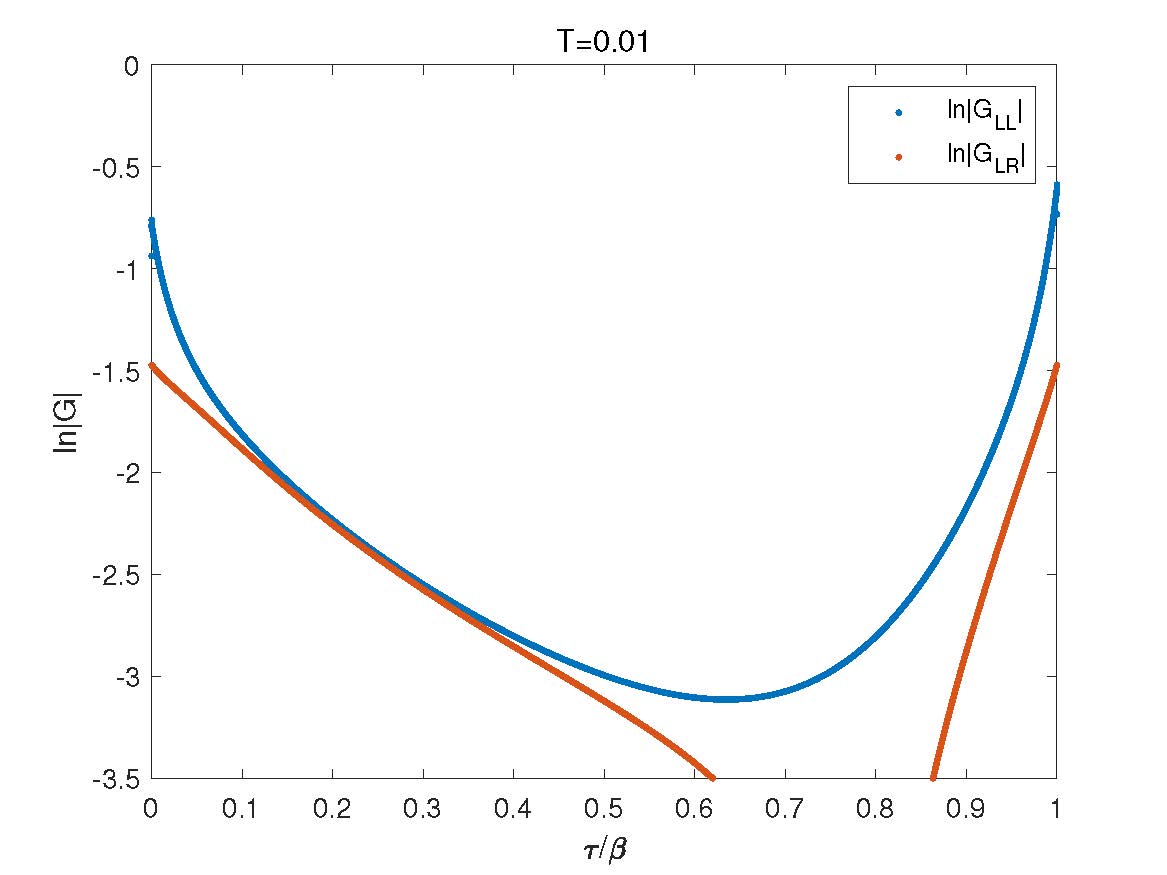}}
	\subfigure[]{\includegraphics[scale=.33]{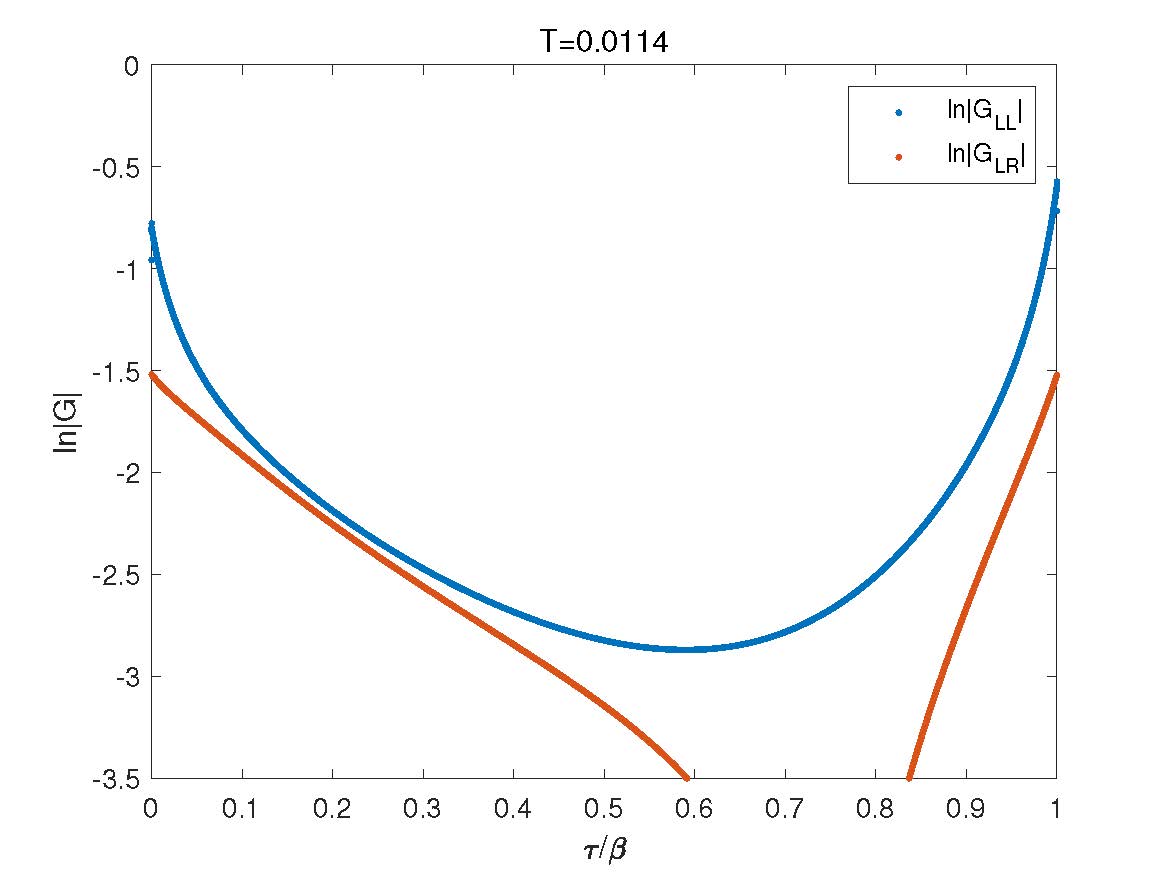}}
	\subfigure[]{\includegraphics[scale=.33]{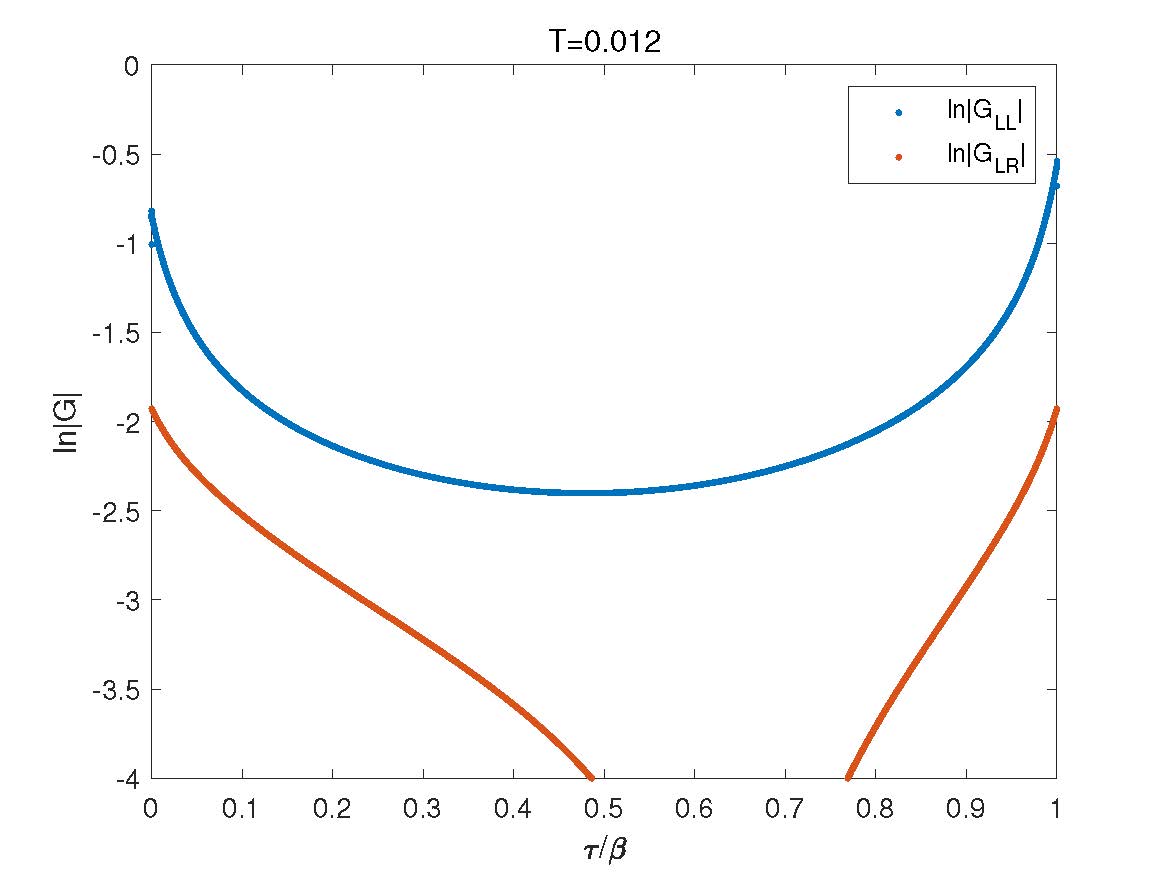}}
	\subfigure[]{\includegraphics[scale=.33]{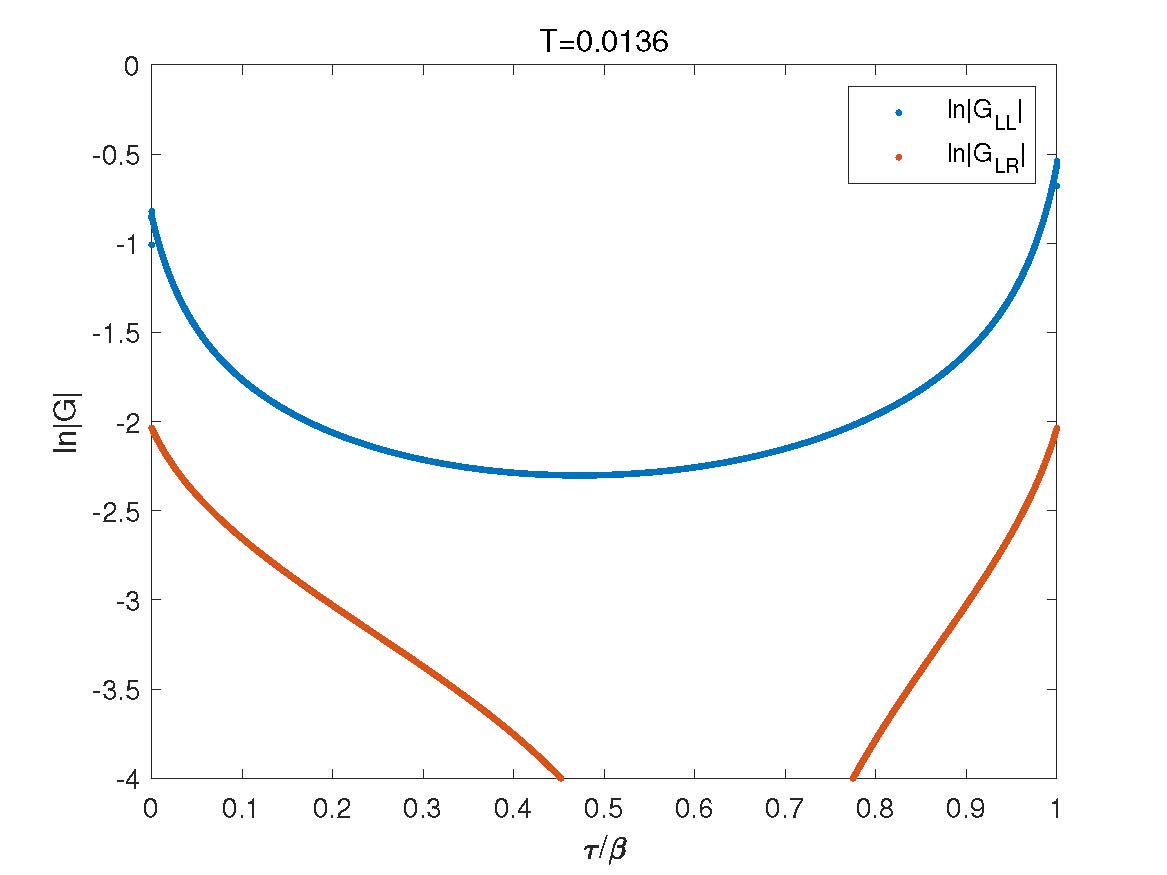}}
	\caption{Left-top: $\Omega(T)$ for $\mu, \kappa$ in the region of two transitions. The other diagrams represent $\ln G$, with $G = G_{LL}, G_{LR}$ versus $\tau/\beta$ in the temperature represented by red circles in  $\Omega(T)$. In the intermediate region, especially close to the low temperature transition, the decay is linear and there is still substantial overlap between $G_{LL}$ and $G_{LR}$ which suggests that some traversability may persist.}\label{fig:lnG}
\end{figure}

\subsubsection{The energy gap $E_g$}
In the case of Majorana fermions, the coupled two-site SYK model \cite{maldacena2018} has a gap $E_g$ in the low temperature limit that is a distinctive feature of the wormhole phase. The almost temperature independence of the grand potential in this limit is a strong indication of a gapped system. A more direct evidence of a gap in the spectrum is directly obtained from the decay of Green's functions. In gapped systems, the decay of $G_{LL}$ or $G_{LR}$ is exponential with a decay rate given by $E_g$. Indeed, we shall see $G_{LL}$ and $G_{LR}$ have a similar exponential decay, indicating a similar probability to stay in each site which implies a continued tunneling between the two sites. This is the type of feature expected in a traversable wormhole.  In the low temperature limit, in the region of parameters where we expect a gap, the Green function is  expressed as,
\begin{equation}\label{expfit}
G_{ab}(\tau)\sim e^{-E_g\tau}f_{ab}(\tau),
\end{equation}
with $a = L,R$ and $b = L,R$.
Numerically, we shall show that, in this limit, and for sufficiently small $\mu$, $f_{LL}(\tau)\approx f_{LR}(\tau)$ are nearly constant. 
We obtain the gap $E_g$ by fitting the numerical $G_{ab}$ to an exponential. This was also one of procedures employed for Majorana fermions Ref.~\cite{maldacena2018} to identify the low temperature phase as a traversable wormhole. In the high temperature limit, we expect a transition, or crossover, to the gapless black hole phase where the ansatz (\ref{expfit}) does not apply. A priori, the nature of the decay is unclear in the intermediate phase of finite but small $\mu$.  

In figure \ref{fig:lnG}, we depict results of the decay of the Green's function in the three phases.  We choose parameters, $\kappa=0.05$ and $\mu=0.072$, where the existence of two transitions in the grand potential, figure \ref{fig:lnG}(a), is more clearly observed . The red circles correspond to the temperatures in which the decay of the Greens's function is studied. As was expected, the Green's function in the low temperature phase, plots (b), (c) in figure \ref{fig:lnG}, decays exponentially and both Green's function have a strong overlap. This is the behavior for Majoranas in the wormhole phase \cite{maldacena2018}. In the high temperature phase, the decays is not exponential for $G_{LR}$ and there is virtually no overlap with $G_{LL}$. This is consistent with a two black hole phase with an explicit coupling between them that does not change the gravitational background.

For intermediate temperatures, between the two phase transitions, the situation becomes more complicated. There is still substantial overlap between $G_{LL}$ and $G_{LR}$, especially close to the low temperature transition. Moreover, the decay is still exponential though in a more limited range of imaginary times which decreases as temperature increases. We tend to believe it is still a wormhole phase but with a worse traversability and, according to previous results, with a finite charge $Q > 0$ and constant entropy in this region. However, further calculations are needed to settle the nature of this intermediate region.

In order to gain more explicit information of the gap $E_g$, we carry out a fitting of the Green's function by an exponential in a broad range of parameters. Results depicted in  the left column of figure \ref{fig:transition}  show that for sufficiently small $\mu$ and not too large $\kappa$, $E_g$ vanishes abruptly at the temperature separating the wormhole from the black hole phase in the same way as in the Majorana case. Only one transition is observed in this region of parameters. As we increase $\mu$, still for small $\kappa$, we access the region where two transitions occur. As was mentioned earlier, the fitting in the intermediate phase becomes less reliable. However, we still observe a sharp drop in $E_g$ but it does not vanishes at the transition. In the intermediate phase, it increases with temperature, very much like the charge does, and finally vanishes at the higher temperature transition towards the black hole phase. We stress the fitting necessary to obtain $E_g$ becomes increasingly unreliable as temperature increases. For instance, the abrupt vanishing of $E_g$, see figure \ref{fig:transition}g, for sufficiently large $\kappa$, is not related to a thermodynamic transition as the grand potential only undergoes a sharp crossover at that temperature.

These results in the intermediate region are confirmed in figure \ref{fig:Eg_Q} (left) for a choice of parameters where the separation between the two transitions is larger. 
Taking into account that the fitting is more reliable in the lower temperature limit, the existence of a finite gap for temperature slightly above the first transition suggests that the intermediate phase may still be described by a wormhole geometry with limited traversability and a finite charge and constant entropy. However, we stress this is a tentative explanation, it may also occur that a finite charge make unstable the wormhole phase and the gap in the intermediate phase is not related to a wormhole geometry. However, this intermediate phase is not the high temperature black hole phase in disguise because the grand potential results indicate that this transition occurs at a higher temperature. Therefore the nature of the intermediate phase, if it is not a charged wormhole, would still be an open question.

\begin{figure}
	\includegraphics[scale=.3]{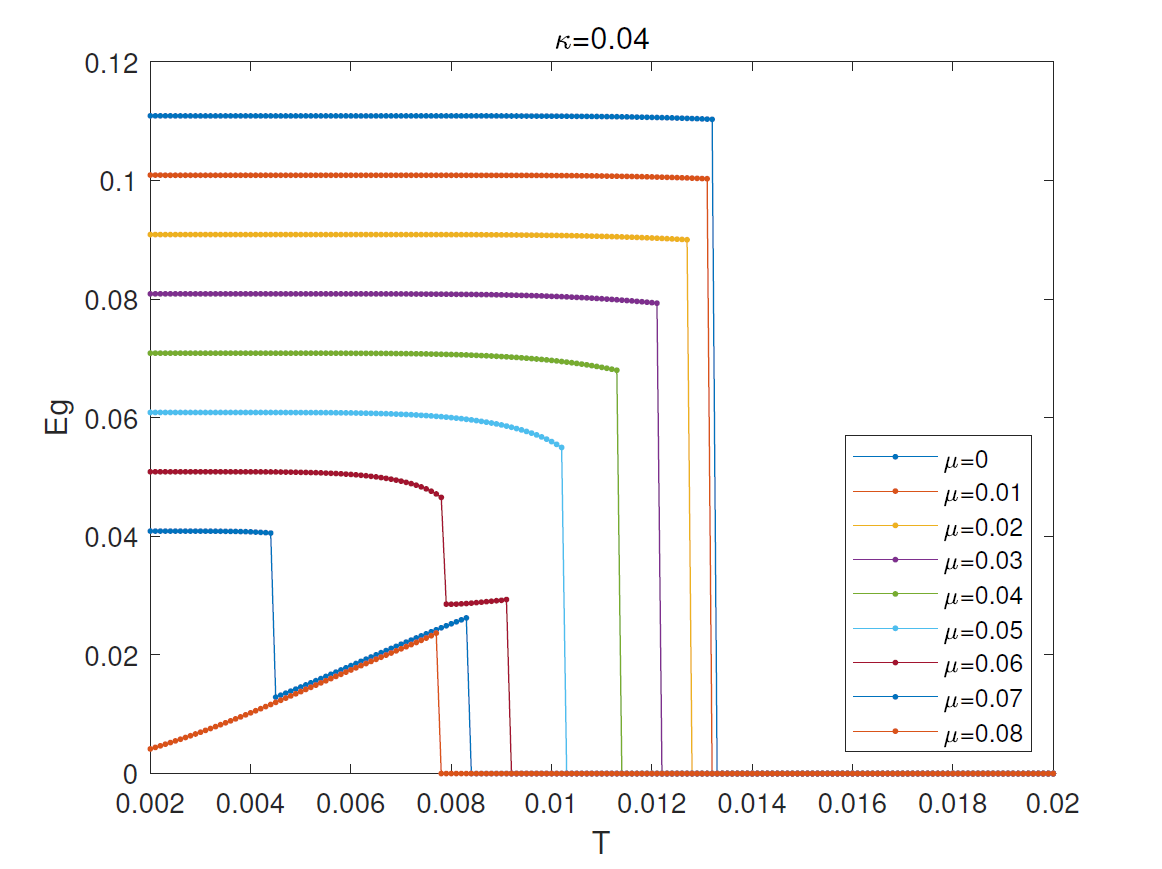}
	\includegraphics[scale=.3]{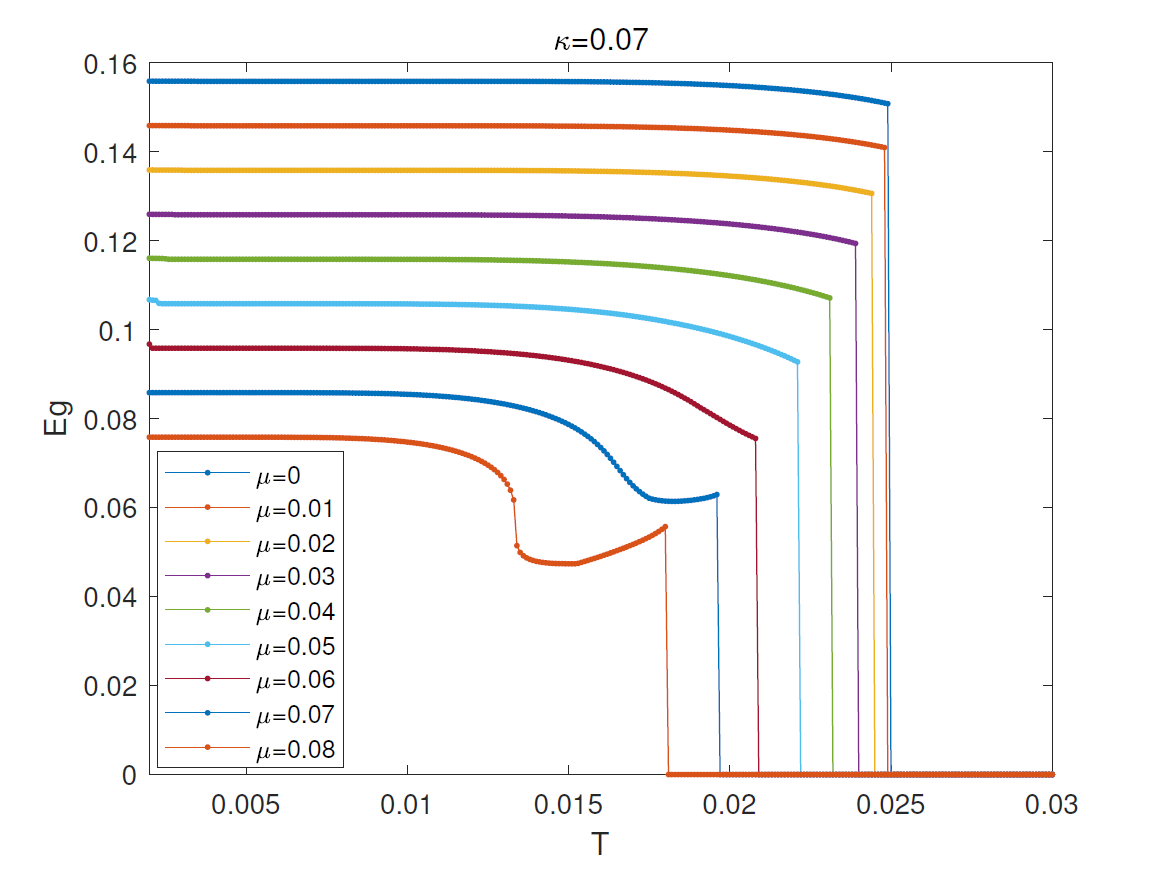}
	\caption{Summary of the gap $E_g$ as a function of $T$ in the region of parameters for which two phase transitions coexist. }\label{fig:Eg_kp_05_07mu_0_09}
\end{figure}

Clearly, further calculations are needed to reach a firm conclusion. For the moment, we conclude this section with a summary, see figure \ref{fig:Eg_kp_05_07mu_0_09}, of the detailed dependence of $E_g(T)$ with $\kappa$ and $\mu$ on a broader range of parameters that includes the intermediate region between the two transitions. Results are fully consistent with previous findings for the charge and grand potential.

\subsection{Phase diagram}
We now combine and extend previous results in order to provide a rather comprehensive description of the system's phase diagram for a given coupling $\kappa$, temperature $T$ and chemical potential $\mu$.

\begin{figure}
	\subfigure[]{\includegraphics[scale=.35]{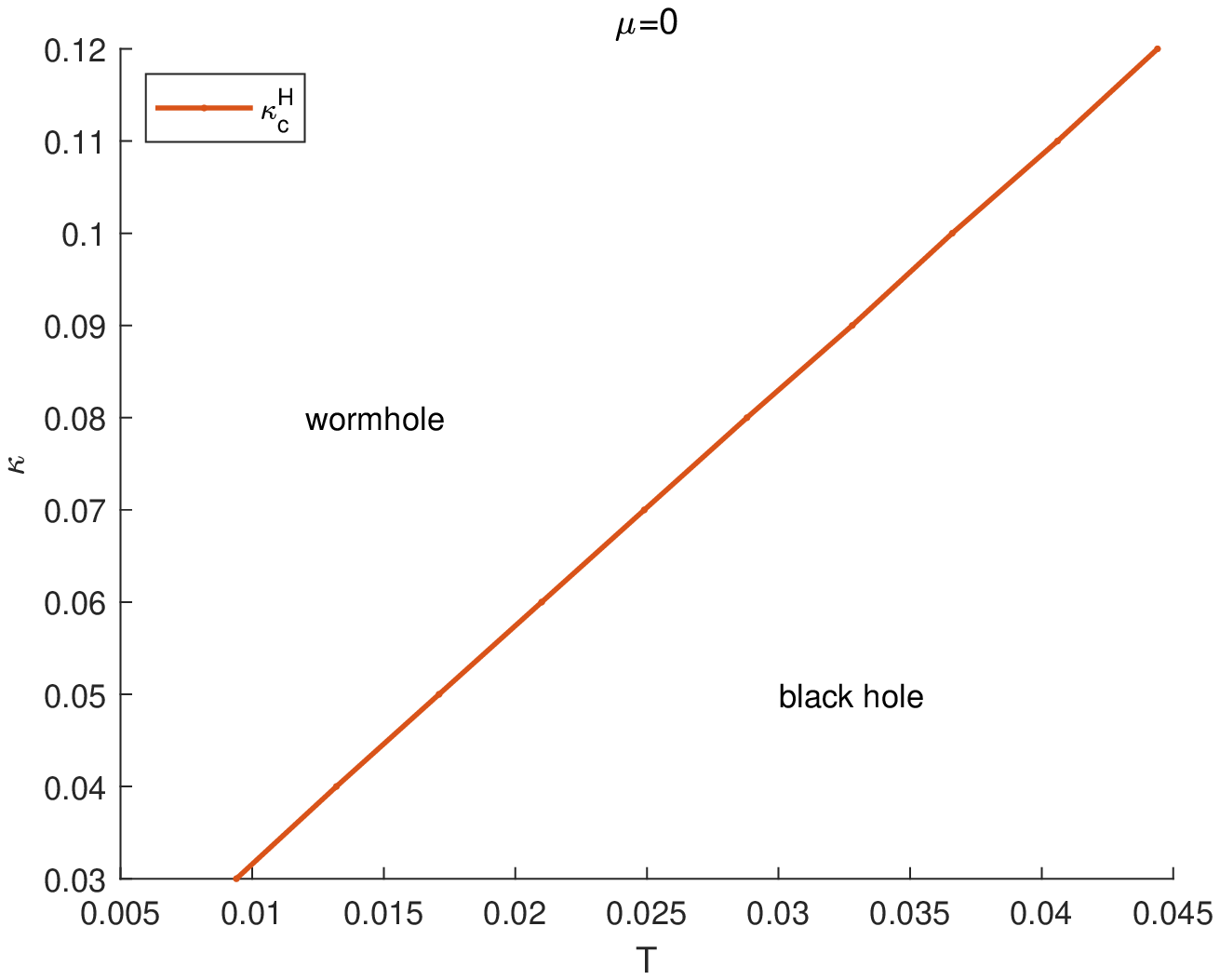}}
	\subfigure[]{\includegraphics[scale=.35]{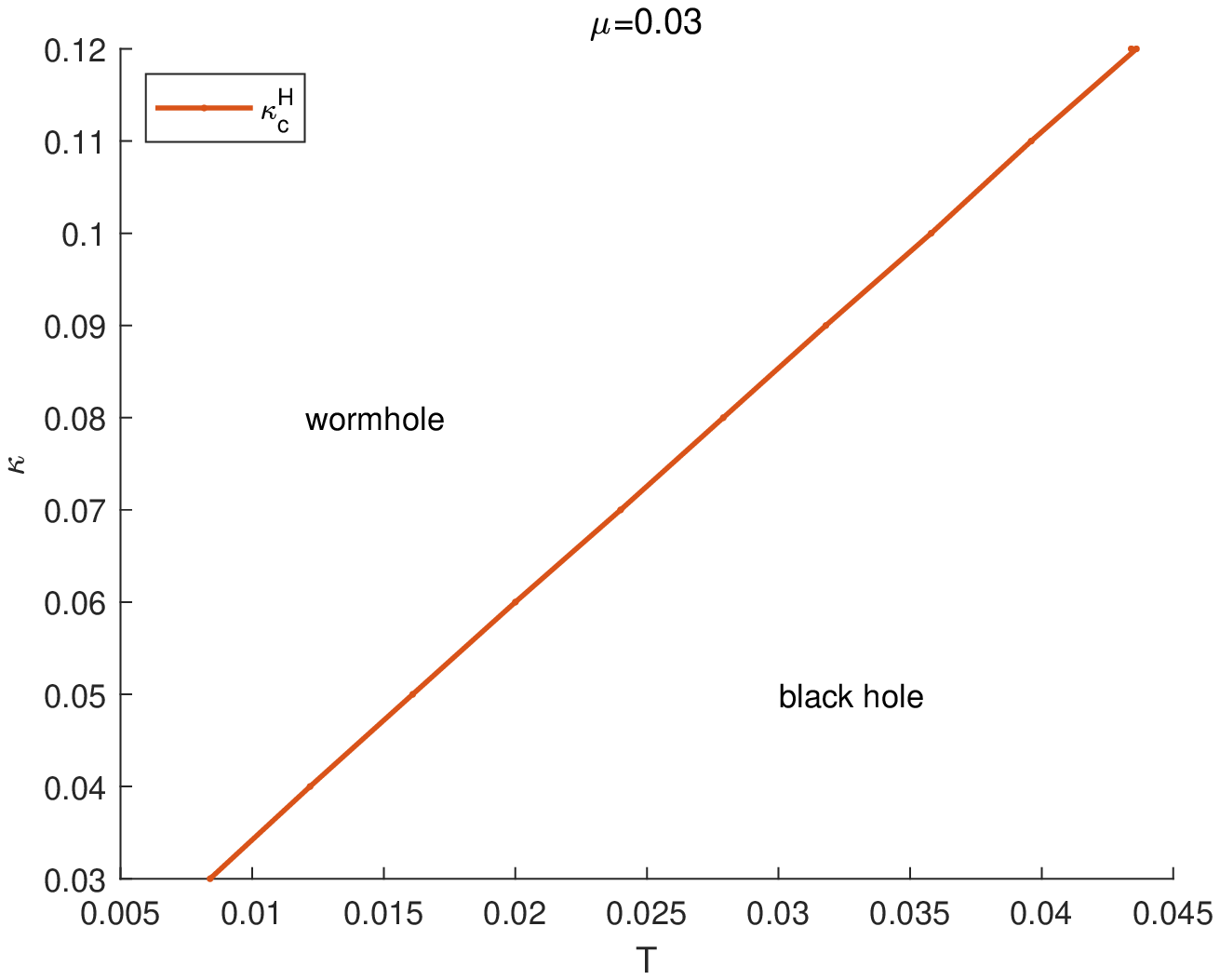}}
	\subfigure[]{\includegraphics[scale=.35]{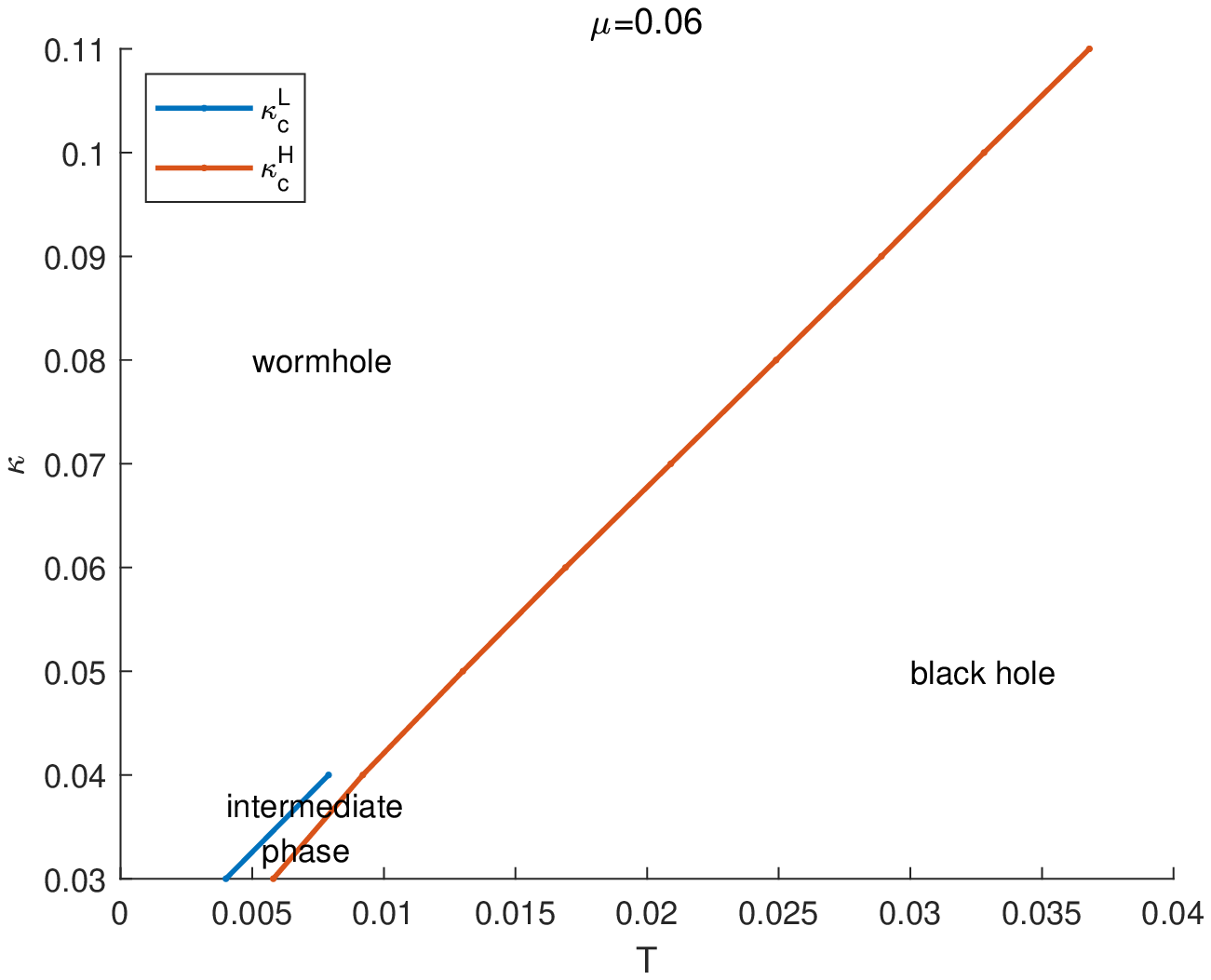}}
	\subfigure[]{\includegraphics[scale=.35]{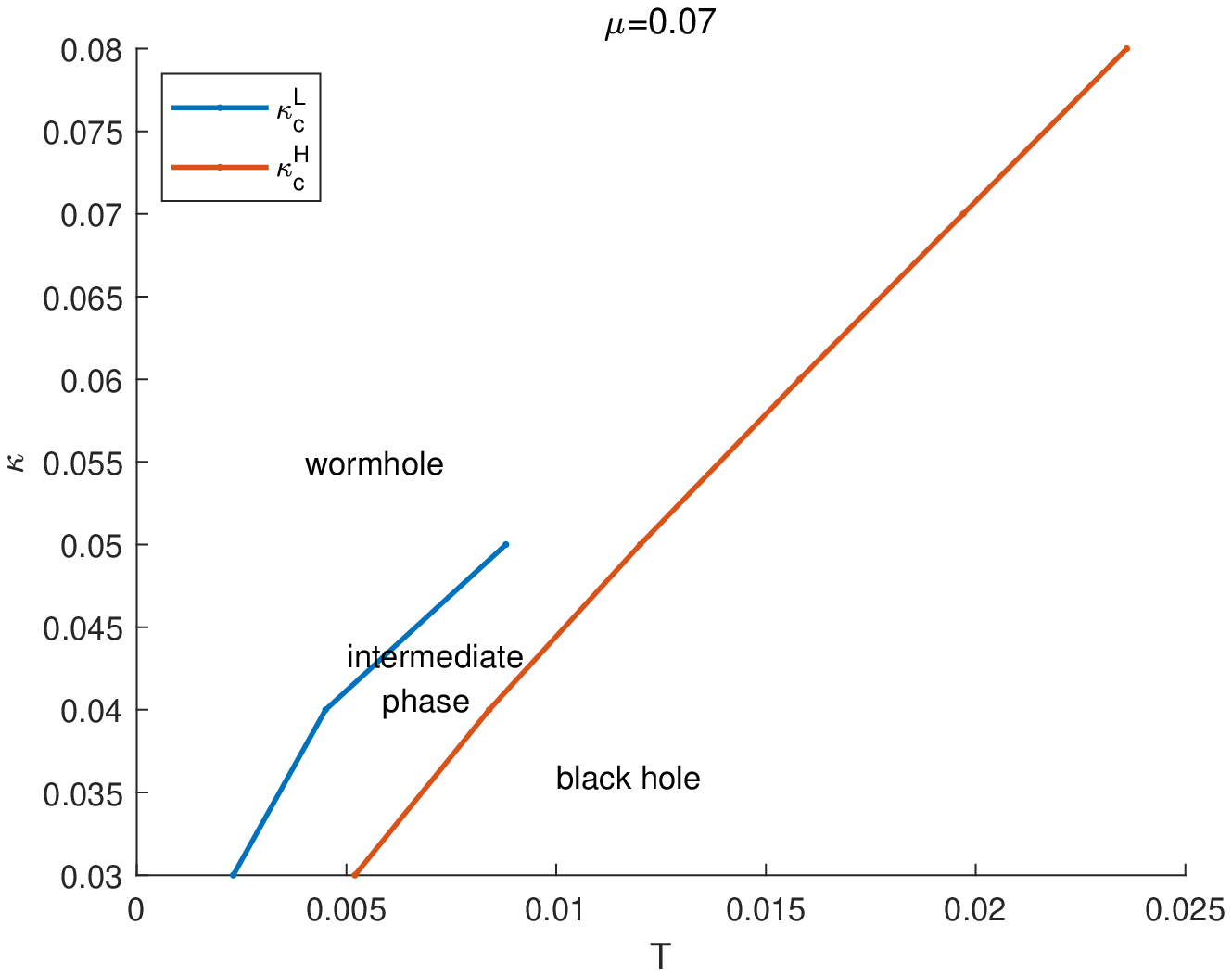}}
	\subfigure[]{\includegraphics[scale=.35]{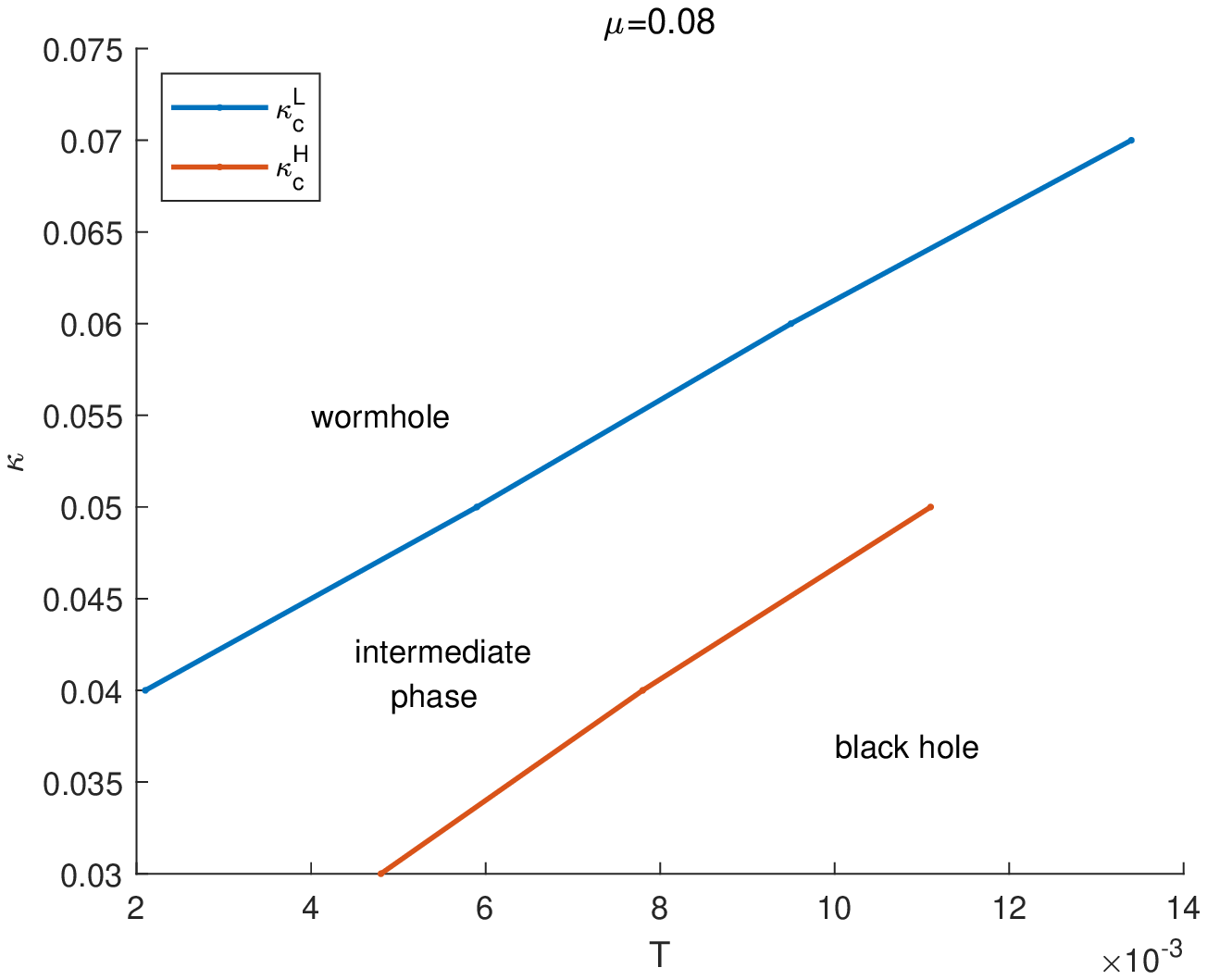}}
	\subfigure[]{\includegraphics[scale=.35]{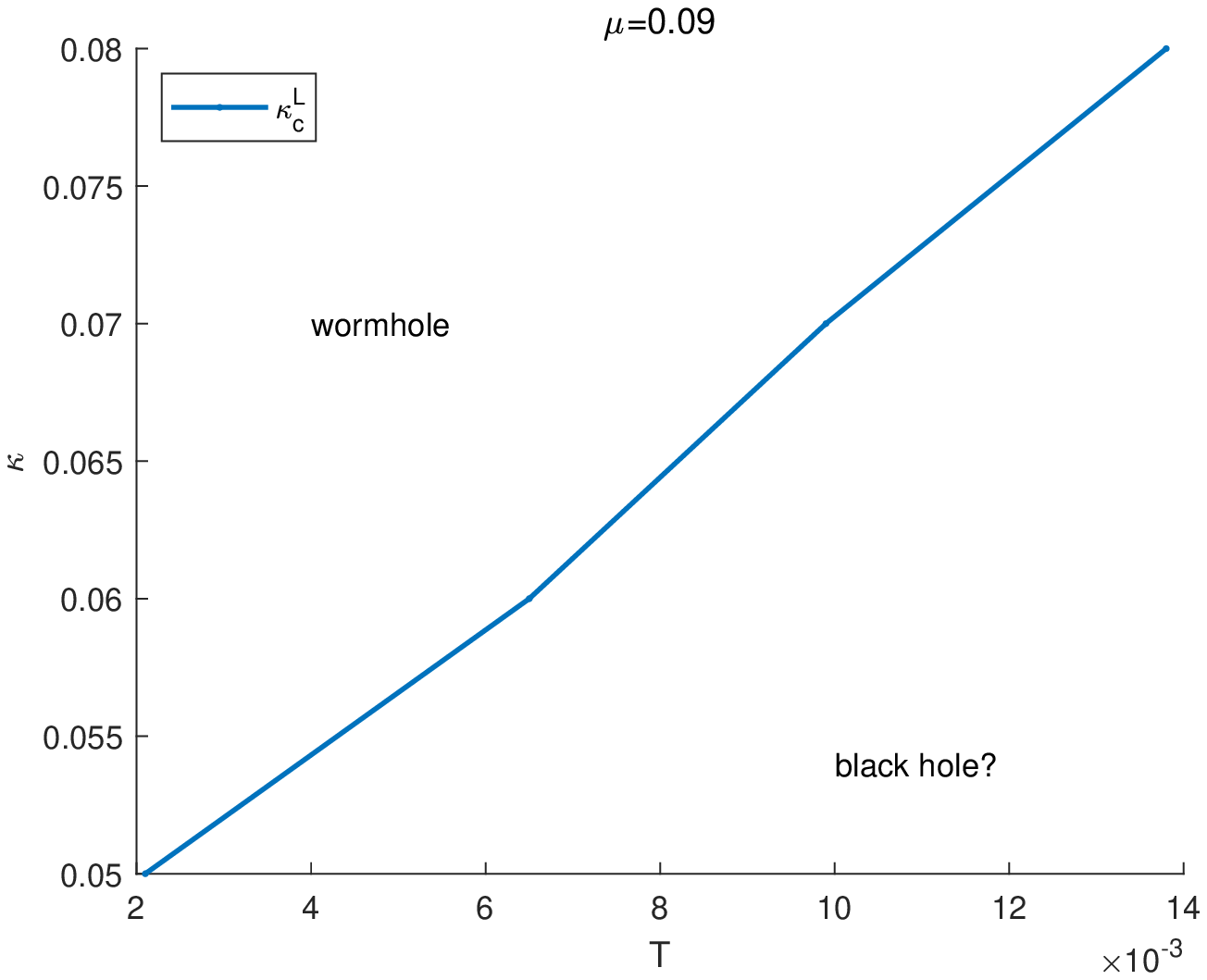}}
	\caption{Phase diagram of the model as a function of the coupling $\kappa$ and temperature $T$ for different $\mu$'s obtained from the singularities of the grand potential.
		Blue (red) lines represent the critical $\kappa$ corresponding to the low (high) temperature phase transition. The end of lines indicate that the phase transition is replaced by a crossover. The intermediate phase, that may correspond to a charged wormhole, is only observed in a limited range of parameters. }\label{fig:PT_ka_p15_mu_09}
\end{figure}

The phase diagram for various $\mu$'s, extracted from the grand potential, are shown in figure \ref{fig:PT_ka_p15_mu_09} as a function of $\kappa$ and temperature. Lines indicate first order phase transitions and their ends signal that the transition becomes a crossover. For sufficiently small $\mu$, the phase diagram is essentially identical to that of two coupled Majoranas SYK's where there is a first order phase transition between the wormhole and the black hole phase that becomes a crossover for large $\kappa$.

As $\mu$ increases, the transition occurs at lower temperature. For $\mu\geq 0.06$, and a not too large $\kappa$, we observe two transitions. The one at high temperature corresponds to the transition to the black hole phase. However, there is a new one at lower temperatures which is rather unexpected. Based on the previous results for the charge and the gap, it could in principle be a transition from the wormhole to a charged AdS background with no traversability or to a charged  traversable wormhole with limited traversability. As was mentioned earlier, results for the energy gap $E_g$ suggest the latter but more information is needed to clarify this issue. As $\mu$, or $\kappa$ increases further, we shall see the two transitions become eventually crossovers and this intermediate region ceases to exist.    

\begin{figure}
	\includegraphics[scale=.54]{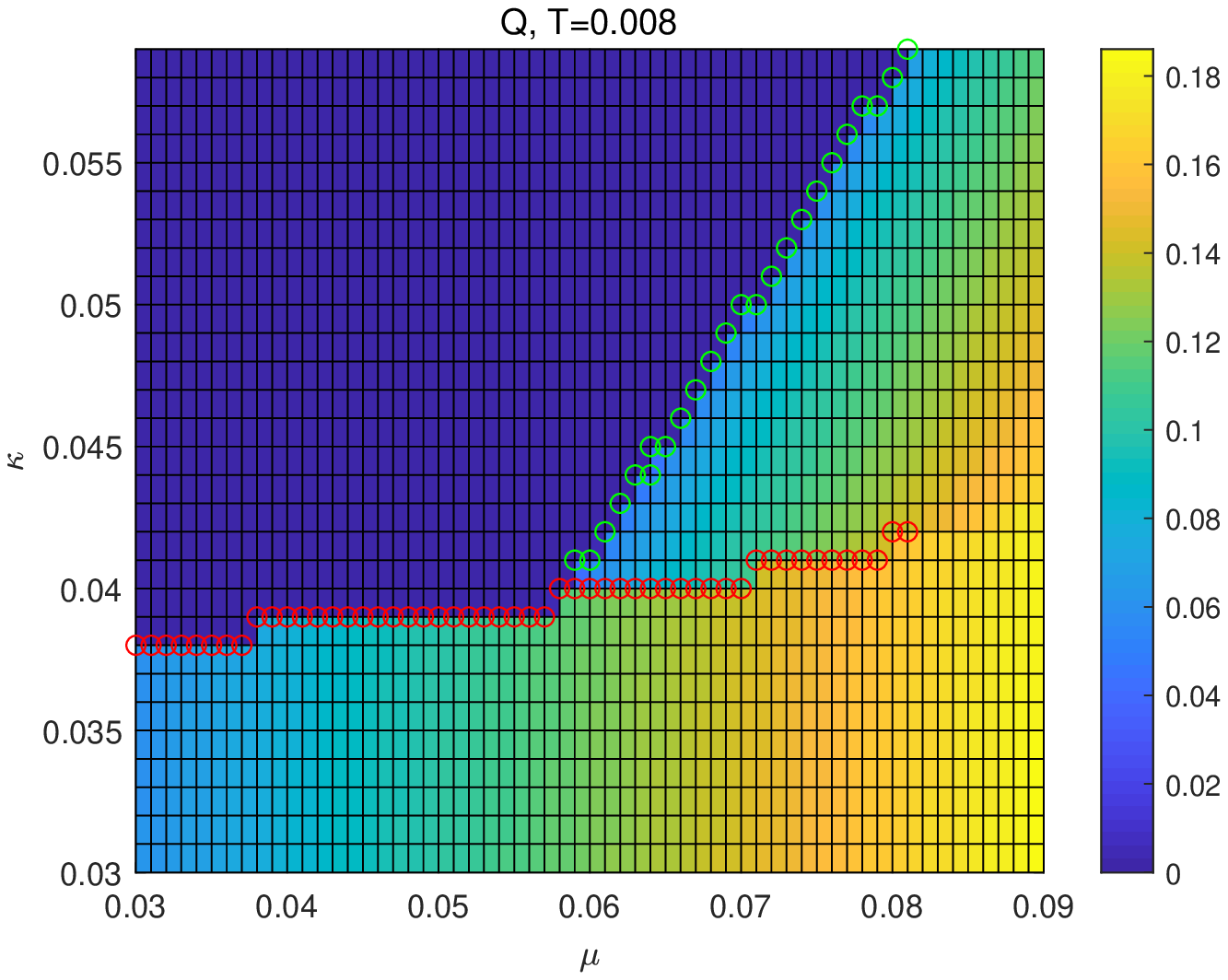}
	\includegraphics[scale=.54]{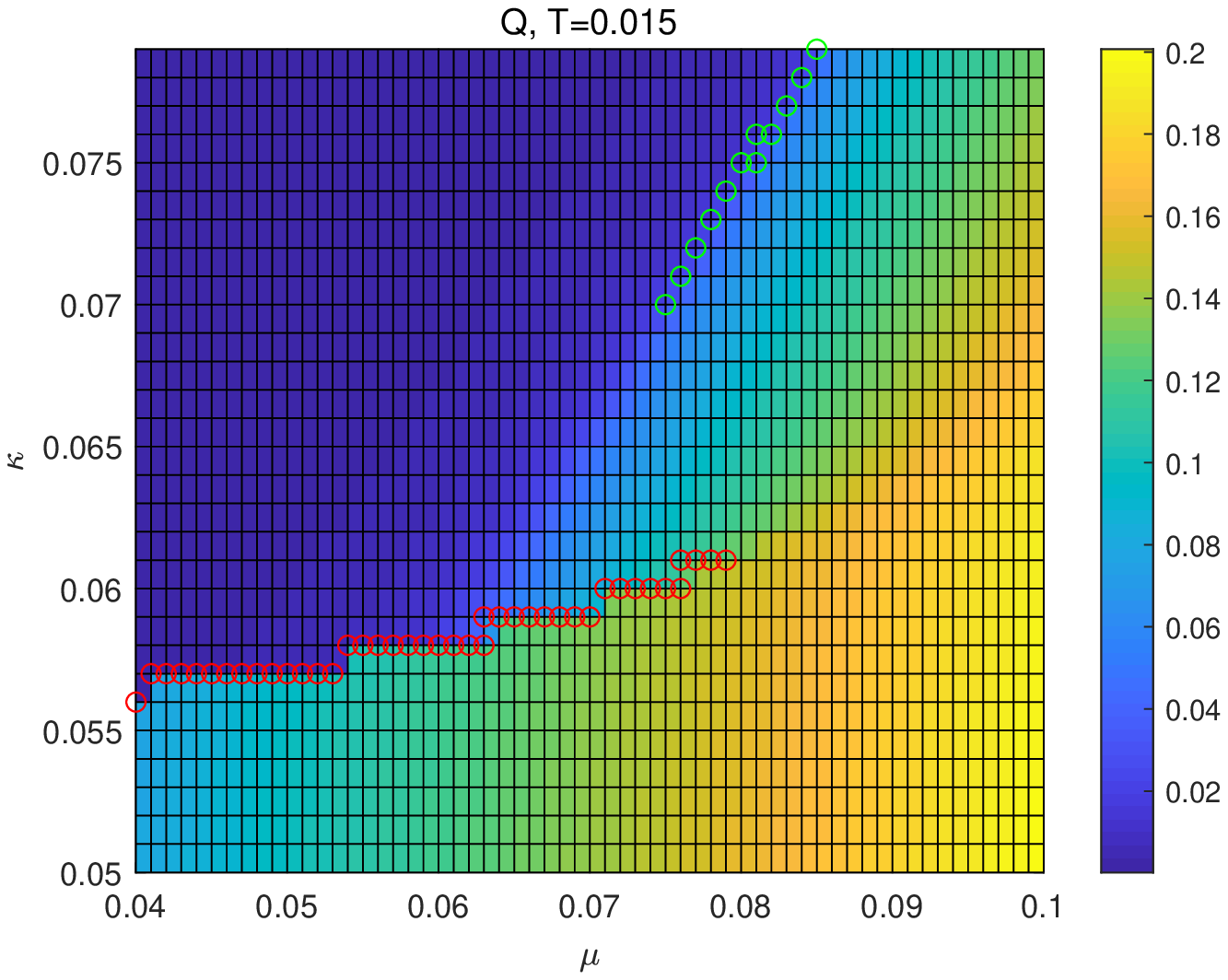}
	\caption{Charge $Q$ in the $\kappa-\mu$ plane for $T=0.008$(left) and $T=0.015$(right). The red circles denote the high temperature wormhole to black hole phase transition, while the green circles denote the novel low temperature phase transition liklely between the cold wormhole phase and a phase of limited traversability that may still correspond with a wormhole though with finite charge. Left: Temperature is low enough for the intermediate phase between the two phase transition (light-blue to green region) to be observed. Right: Same but for a higher temperature where the low temperature phase transition has turned in a crossover and only the high temperature transition persists.} \label{fig:Q_ka_mu}
\end{figure}

In order to gain a further understanding of the phase diagram, we compute the charge $Q$ 
in the $\kappa-\mu$ parameter space. As is observed in figure \ref{fig:Q_ka_mu}, in the low-temperature (left figure), small-$\mu$ limit, the dependence of $\mu$ is very weak and only one transition is  observed, denoted by red circles, between a wormhole phase for larger $\kappa$, and a black hole for lower $\kappa$. There is an intermediate region in $\kappa-\mu$ space where the two transitions mentioned above are observed. We note that the one for larger $\kappa$ corresponds to a wormhole phase. The intermediate phase between the two transitions has a finite charge and, as shown earlier, the gap $E_g$ is still finite so it may still a wormhole but charged and at finite temperature. For sufficiently large $\mu$ the black hole transition becomes a crossover. At sufficiently higher temperature (right plot), the low temperature phase transition will become crossover in the region of $\mu$ where the high temperature phase transition happens. Therefore we only observe a transition to a charged black hole. This is fully consistent with the phase diagram obtained from the grand potential.

\begin{figure}
	\includegraphics[scale=.7]{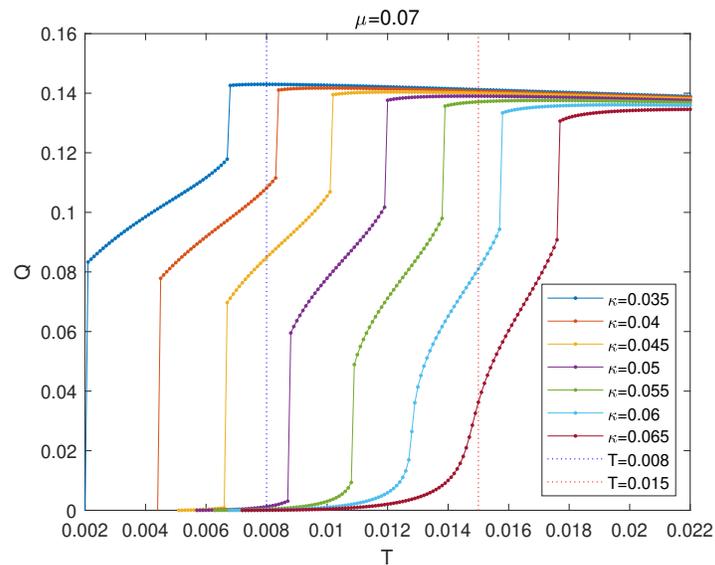}
	\caption{$Q(T)$ for $\mu=0.07$ with $\kappa=0.035,\cdots,0.065$. For small $\kappa$, both transitions, characterized by jumps in the charge, are observed. However, as $\kappa$ increases, eventually the low temperature transition becomes a crossover. The vertical lines correspond with the temperatures employed in figure \ref{fig:Q_ka_mu}. As was expected, results of both figures are fully consistent.} \label{fig:Q_mu_07}
\end{figure}

This picture of the intermediate phase is confirmed explicitly in figure \ref{fig:Q_mu_07} for $\mu=0.07$. A fixed and sufficiently low temperature $T = 0.008$, represented by a blue dotted
vertical line on the left, will intersect the charge curve at couplings $\kappa$ belonging to the black hole, the intermediate and the wormhole phase in agreement with the results of figure \ref{fig:Q_ka_mu}. 
Likewise, a red dotted vertical line for a higher temperature will intersect the charge for couplings $\kappa$ that, according to figure \ref{fig:Q_ka_mu}, belong to either the crossover between the chargeless and possibly the charged wormhole or the black hole phase which confirms that only one transition exists. Eventually, as we further increase $\kappa$, we will see a similar behavior in the high temperature phase transition towards a black hole which will become a crossover. 

\begin{figure}
	\includegraphics[scale=.33]{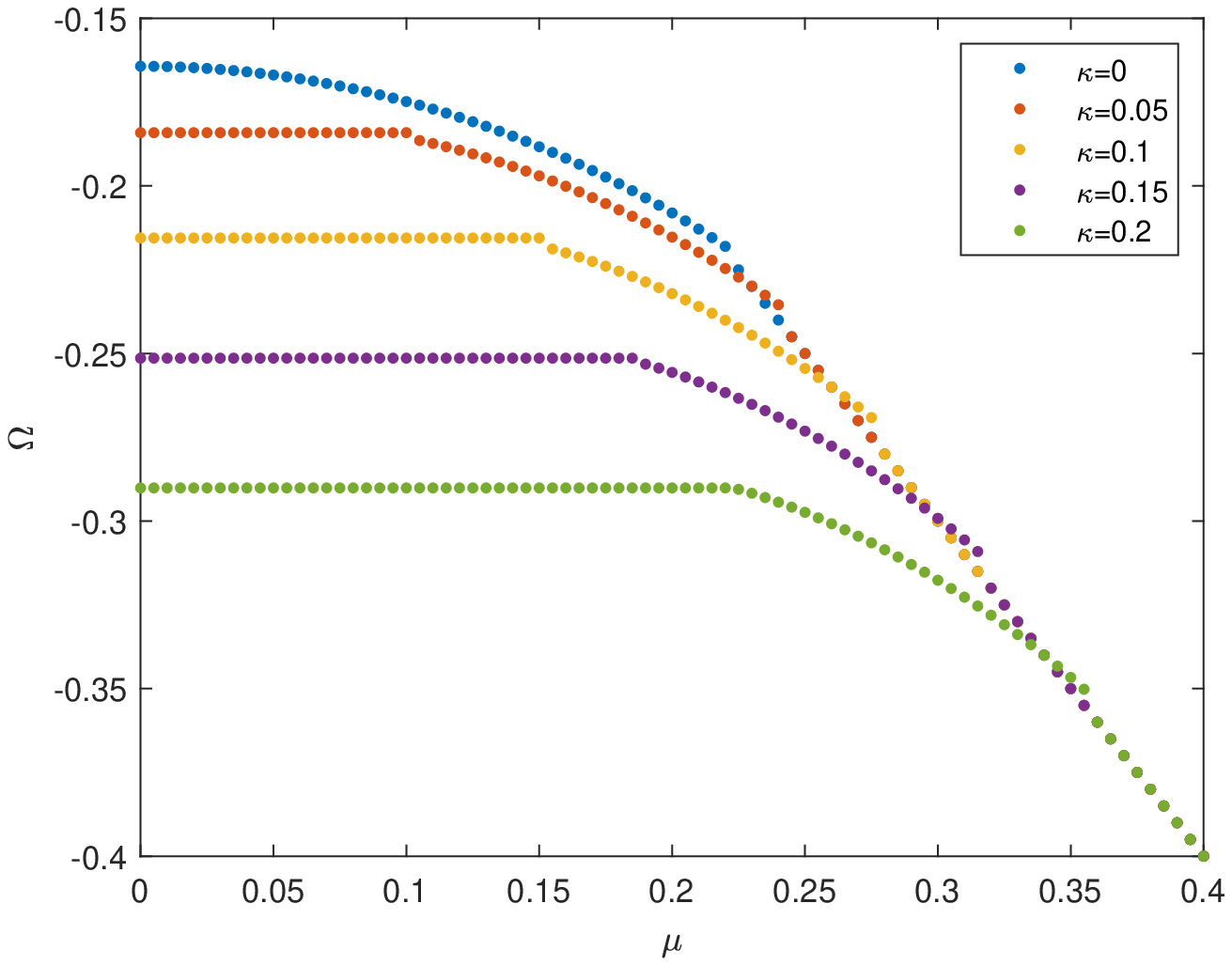}
	\includegraphics[scale=.33]{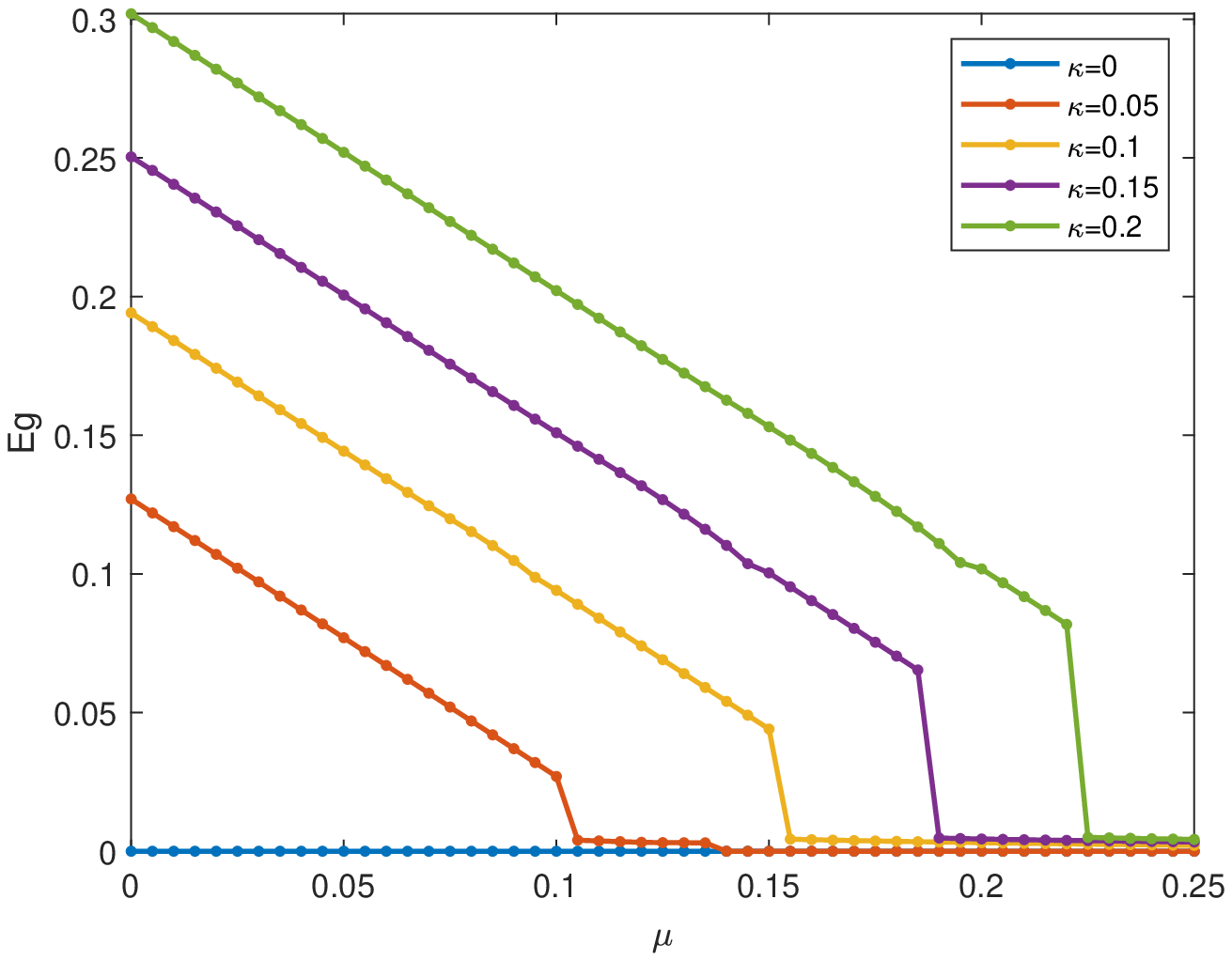}
	\includegraphics[scale=.33]{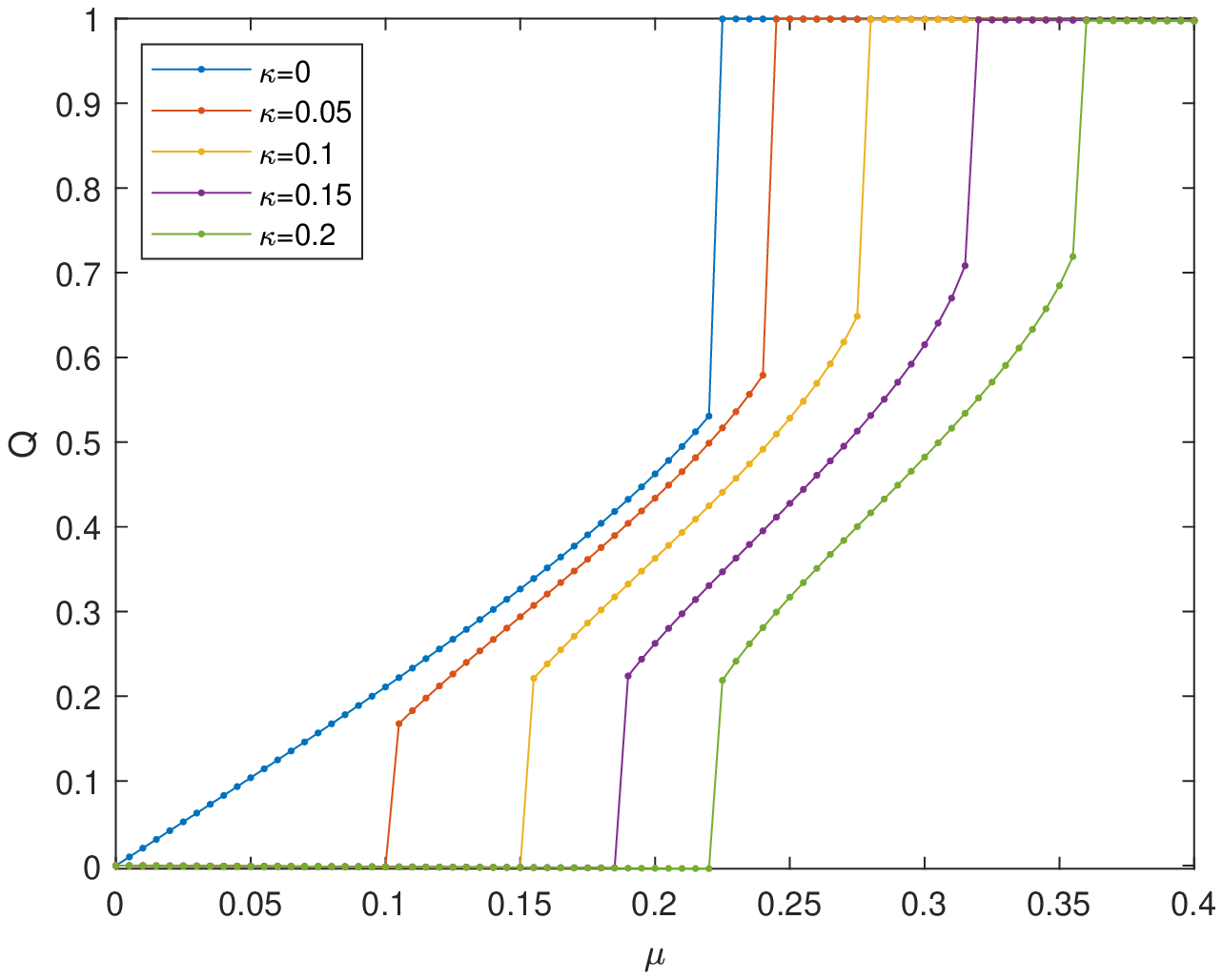}
	\caption{The grand potential $\Omega$, the charge $Q$ and the gap $E_g$ in the low temperature limit for a broad range of parameters $\mu$ and $\kappa$. The transition in the charge for large $Q\geq 0.2$ is not related to the wormhole phase as it also occurs in the single complex SYK \cite{sorokhaibam2020}. $E_g$ decreases monotonically with the chemical potential. }\label{fig:Eg_klowT}
\end{figure}

Finally, for the sake of completeness, we study, see figure \ref{fig:Eg_klowT}, the gap, the charge and the grand potential, for larger values of $\kappa$ and $\mu$ and the lowest temperature that we can reach numerically. 
Interestingly, we observe $E_g$ decreases monotonically and almost linearly with $\mu$, for a fixed low $T$ and different values of $\kappa$. It eventually vanishes abruptly for sufficiently large $\mu(\kappa) \sim 0.2$ and it is zero for $\kappa =0 $ which suggests $E_g$ is related with the wormhole phase and that $\mu$ is only a shift in energy that reduces its value. This is confirmed by the dependence on $\mu$ of the grand potential for $\kappa = 0$. This is the only case where $\Omega$ is not completely flat for small $\mu$ which is an indication that $E_g = 0$ in this case. For finite $\kappa$, we observe a transition for small $\kappa$ and a crossover for larger $\kappa$.

Another interesting feature is the observed jump in the charge $Q$ for large values of $\mu > 0.2$, which are much larger than those corresponding to the transitions studied earlier. Similar results have been reported \cite{sorokhaibam2020} in single complex SYK so we do not think that it has any relation with the wormhole phase.

\section{Low energy effective action}\label{sec:lowenergy}
In this section, we exploit symmetries of the SD equations in the infrared limit in order to find out the low energy effective action of the model.

At low energies $\omega,T\ll J$ the SD equations \eqref{eq:c_SDequations} can be written compactly as:
\begin{subequations}\label{eq:low_energy_SD}
	\begin{align}
	&\tilde{\Sigma}_{ab}(\tau,\tau') = -\left(-1\right)^{q/2} J^2 s_{ab} \left[G_{ab}(\tau,\tau')\right]^{q/2} \left[G_{ba}(\tau',\tau)\right]^{q/2-1} -\bm{\eta}_{ab} \delta(\tau-\tau')\,,\label{eq:low_energy_SD1}\\
	&\sum_{b} \left[G_{ab}\star \tilde{\Sigma}_{bc} \right]=- \delta_{ac} \delta(\tau-\tau') \,,\label{eq:low_energy_SD2}
	\end{align}
\end{subequations}
where 
\begin{align}
\bm{\eta}_{ab}=\begin{pmatrix}
-\partial_{\tau} + \mu&\eta\\
\eta^{*}&-\partial_{\tau} + \mu
\end{pmatrix}\,. 
\end{align}
Ignoring the $\bm{\eta}_{ab}$ terms in \eqref{eq:low_energy_SD1}, the above system of SD equations \eqref{eq:low_energy_SD} possesses the following time reparametrization and $U(1)$ gauge symmetries:
\begin{subequations}\label{eq:c_repu1}
	\begin{align}
	G_{ab}(\tau,\tau') &\to \left[f'_{a}(\tau)\,f'_{b}(\tau')\right]^{\Delta} e^{i \left(\Lambda_{a}(\tau) - \Lambda_{b}(\tau')\right)} G_{ab}\left(f'_{a}(\tau)\,f'_{b}(\tau')\right)\,,\label{eq:c_repu1_1}\\
	\tilde{\Sigma}_{ab}(\tau,\tau') &\to \left[f'_{a}(\tau)\,f'_{b}(\tau')\right]^{1-\Delta} e^{i \left(\Lambda_{a}(\tau) - \Lambda_{b}(\tau')\right)} \tilde{\Sigma}_{ab}\left(f'_{a}(\tau)\,f'_{b}(\tau')\right) \label{eq:c_repu1_2}\,,
	\end{align}
\end{subequations}
where $f_a\in \textrm{Diff}(S^1),\,\Lambda_a\sim \Lambda_a +2\pi$, and the winding number $n_a$ of the compact gauge parameter $\Lambda_a$ is conjugate to the $U(1)$ charge $Q_a$. 

\subsubsection{High Temperature}
Let us first discuss the high temperature limit where we expect that the coupling $\eta$ can be neglected and the SD equations \eqref{eq:c_SDequations} can be solved by an ansatz in which all $L-R,R-L$ functions vanish, describing two copies of a complex SYK model in a thermal state dual to black hole. Standard arguments \cite{maldacena2016} imply that, at low energies, the effective action of the system is 
\begin{align}\label{eq:coupled_low_energy_action_th}
S = Schw[f_{L},\Lambda_{L}] + Schw[f_{R},\Lambda_{R}]\,,
\end{align} 
where the Schwarzian of a single complex SYK model can be expressed in terms of the time reparametrization and $U(1)$ gauge symmetries in \eqref{eq:c_repu1} \cite{davison2017,Gu:2019jub}
\begin{align}\label{eq:Schwa}
Schw[f_{a},\Lambda_{a}] = - N\alpha_{S} \int d\tau \left\lbrace \tan \frac{f_{a}(\tau)}{2},\tau \right\rbrace  +\frac{NK}{2} \int d\tau \left( \Lambda'_{a}(\tau) +i \mathcal{E}_{a} f'_{a}(\tau) \right)^2 \,,
\end{align}
following the notation of \cite{Gu:2019jub}. Here $\mathcal{E}_{a}$ is related to the charge $Q_{a}$ with $a = L, R$. The coefficient $\alpha_{S}$ is the prefactor of the heat capacity in the low temperature limit and $K$ is the compressibility. Both are of the order of $J^{-1}$, which from here on we set to $1$. 

Each copy of the complex SYK model has its own global $SL(2)\times U(1)$ symmetry \cite{davison2017},
\begin{align}\label{eq:gl_symmetry_un}
\delta f_a=\epsilon^0_a +\epsilon^{2}_a e^{i f_a} +\epsilon^{3}_a e^{-i f_a}\,,\qquad \delta \Lambda_{a}=-i\mathcal{E}\,\delta f_{a} +\epsilon_a\,,
\end{align}
which should be treated as a gauge symmetry, since it leaves the bilocal fields $G_{ab},\tilde{\Sigma}_{ab}$ in \eqref{eq:c_repu1} invariant.

The thermodynamics in this phase is (twice) the usual complex SYK thermodynamics. At temperature $\beta$, the grand potential takes the form \cite{gu2020},
\begin{align}\label{eq:grand_potential_thermal}
\frac{\Omega}{2N} = f(\mu_0) -\mathcal{G}(\mathcal{E})\, \beta^{-1} -2\pi^2\alpha_{S}\, \beta^{-2}\,,
\end{align}
where we have assumed that the two systems are identical. In the above, $f$ is the ground state energy, $\mu_0=\mu+2\pi\,\mathcal{E}\,\beta$, and $\mathcal{G}(\mathcal{E})$ is the Legendre transform of the entropy $\mathcal{G}(\mathcal{E})=\mathcal{S}(\mathcal{E}) -2\pi\mathcal{E} Q$.
The explicit analytical expressions for $\mathcal{\mathcal{E}}$ and $Q$ in the case of the single complex SYK model can be found in Ref.~\cite{gu2020}.

Including small $\eta$ corrections, this expression is also expected to describe the high temperature phase of the coupled system. These will induce $L-R,R-L$ correlators of the order of $\eta$, i.e. $\Delta G_{LR}\sim\eta\, G_{LL}\star G_{RR}$, and thus corrections to the grand potential \eqref{eq:grand_potential_thermal} as well.

\subsubsection{Low Temperature}
In the low temperature limit, we expect that the ground state is related to the traversable wormhole. 
In the Majorana case, it was argued that, in this region, the ground state of the coupled system is close to the TFD state at a particular fictitious temperature $\tilde{\beta}(\eta)$ \cite{maldacena2018,garcia2019} that depends on coupling $\eta$ between sites.  We note that the physical temperature is completely unrelated to $\tilde{\beta}(\eta)$.

Intuitively, a $\tilde{\beta}=\infty$ TFD state is the ground state of the system when $\eta=0$, while the $\tilde{\beta}=0$ TFD state is the ground state when $\eta \to \infty$. Therefore, the only IR effect of the UV parameter $\eta$ is to tune the value of the fictitious temperature $\tilde{\beta}$. The $L-R,R-L$ Greens's functions can be obtained simply by analytically continuing the single-sided correlators to complexified time with imaginary part $\tilde{\beta}/2$, as reviewed in \cite{Fidkowski:2003nf}.\footnote{From the dual gravity perspective this is reflected in the fact that in the complexified maximally extended black hole spacetimes, the time on the left part of the wormhole has an imaginary part $\tilde{\beta}/2$.}  Additionally, for small enough $\eta$, we can assume that we are in the conformal regime of the SYK models, in which case the low energy effective action will be given by the ``valley'' of pseudo-Goldstone modes, plus the interaction potential evaluated on this valley. This leads to a new ground state, but still on the valley.

Note that in a single complex SYK model, the chemical potential $\mu$, which is also a UV parameter, enters the low energy Green's functions $G(\tau)$ through an IR parameter $\mathcal{E}$.\footnote{Alternatively, $\mathcal{E}$ can be more naturally related to the charge $Q$ instead of the chemical potential $\mu$. Note that the Legendre transform which takes us from the canonical to the grand canonical ensemble is equivalent to a Fourier transform of the thermal partition function \cite{Hawking:1995ap,Braden:1990hw}.} In particular, for the single complex SYK model, a consistent extrapolation from the UV exists, determining $\mathcal{E}$ as a smooth, odd function of $\mu$ as long as $|\mu|\lesssim0.24$ \cite{Gu:2019jub}. 

It is then natural to combine the above arguments and assume that the ground state of the coupled complex SYK systems with total Hamiltonian $H_{total}=H_L+H_R+H_{int}$ (\ref{hami}), is close to the charged TFD state \cite{Andrade:2013rra,iqbal2015}. Based on this assumption,  we can use conformal field theory predictions, also for left-right correlations, together with the effective infrared symmetries of the SD equations mentioned earlier, to compute the Green's functions of our system in the infrarred limit,\footnote{Here we are conventionally setting the period of the thermal circle to $2\pi$, i.e. $\tau\sim\tau+2\pi$. After reparametrizing $\tau\to f(\tau)$, the period will be given by the $\tilde{\beta}$ mentioned below.} 
	\begin{align}
	G_{LL}(\tau) &=G_{RR}(\tau) =-b^{\Delta}\frac{e^{\mathcal{E}(\pi-\tau)}}{\left[ 2 \sin \frac{\tau}{2} \right]^{2\Delta}} \,,\label{eq:c_IR_saddleLLRR}
	\end{align}
	\begin{align}
	G_{LR}(\tau) &=e^{-2i\theta} G_{RL}(\tau)=-b^{\Delta} e^{-2i\pi\Delta} \frac{e^{-\mathcal{E}\tau}} {\left[ 2 \cos \frac{\tau}{2}\right]^{2\Delta}}\,.\label{eq:c_IR_saddleLRRL}
	\end{align}
where $\Delta=1/q$, the parameter $b$ is $\tau$ independent but may depend on $\mathcal{E}$, $J$ and $\Delta$.
The spectral asymmetry factor $\mathcal{E}$ depends in a non-universal way on the microscopic parameters of the model.

Let us now describe the derivation of low energy effective action, which involves the reparametrizations $f_a$ and gauge transformation $\Lambda_a$ entering \eqref{eq:c_repu1}. We observe that there is a $\left(\textrm{Diff}(S^1)\times LU(1)\right)^2$ emergent symmetry at low energies. This is spontaneously broken by a solution of our choice, such as \eqref{eq:c_IR_saddleLLRR}, \eqref{eq:c_IR_saddleLRRL}, down to such $f_a,\Lambda_{a}$ which leave the solution invariant.

We thus see that we need to consider $f_a,\Lambda_{a}$ modulo the global $SL(2)\times U(1)$ symmetry, whose infinitesimal form is
\begin{align}\label{eq:gl_symmetry}
\delta f_L=\epsilon^0 +\epsilon^{2} e^{i f_L} +\epsilon^{3} e^{-i f_L}\,,\qquad \delta f_R=\epsilon^0 -\epsilon^{2} e^{i f_R} -\epsilon^{3} e^{-i f_R}\,,\qquad \delta \Lambda_{a}= -i\mathcal{E} \,\delta f_{a} +\epsilon^1\,.
\end{align}

We note, the symmetry \eqref{eq:gl_symmetry} is not physical. It is a redundancy of the effective description from the SYK point of view, related to the state we want to prepare \cite{Harlow:2018tqv}. From the gravitational point of view, it corresponds to the isometries of the rigid $AdS_2$ space and the globally conserved $U(1)_{+}$ on the boundaries \cite{Sachdev:2019bjn}. We can imagine reducing a $d+2$-dimensional action down to a $1$-dimensional action on the two boundaries of $AdS_2$ and then, for the full boundary action, there is a single $SL(2)$ symmetry from the isometries of $AdS_2$ and a single $U(1)$ from the bulk $d+2$-dimensional gauge field. In our case, this is also made manifest from the form of the coupling in \eqref{eq:S_int}. Thus it should be treated as a gauge symmetry, and therefore the associated Noether charges should vanish \cite{usfuture}.

The above suggests that the low energy effective action of the coupled model can be written as
\begin{align}\label{eq:coupled_low_energy_action}
S = Schw[f_{L},\Lambda_{L}] + Schw[f_{R},\Lambda_{R}] + S_{int}\,,
\end{align}
with the Schwarzians $Schw[f_{a},\Lambda_{a}]$ as in \eqref{eq:Schwa}, and the interaction term $S_{int}$ results from the expressions \eqref{eq:c_IR_saddleLRRL} after performing general reparametrizations $\tau\to f_a(\tau)$ and gauge transformations by $\Lambda_{a}(\tau)$ \cite{Moitra:2018jqs}:
\begin{align}\label{eq:S_int}
S_{int} &= \int d\tau\,\left[\frac{b\, f'_{L}(\tau) f'_{R}(\tau)}{4 \cos^2 \frac{f_{L}(\tau) - f_{R}(\tau)}{2}}\right]^{\Delta} \left[ \eta e^{i \left(\Lambda_{L}(\tau) - \Lambda_{R}(\tau)\right)-\mathcal{E}\, \left(f_{L}(\tau)-f_{R}(\tau)\right)} + c.c.\right]\,,
\end{align}
with the dynamical fields belonging in the coset $\left(\textrm{Diff}(S^1)\times LU(1)\right)^2/\left(SL(2)\times U(1)\right)$.
We note that the above action should be supplemented by additional constraints related to the fact that SL(2)
Noether charges must be zero \cite{maldacena2018}.

We consider the following ansatz as a solution of the classical equation of motion, 
\begin{align}\label{eq:f_lambda_ansatz}
f_L(\tau)=f_R(\tau)\equiv f(\tau)\,, \qquad \Lambda_L(\tau)=\Lambda_R(\tau) +\textrm{const} \equiv \Lambda(\tau)\,.
\end{align}

The action is then simplified to
\begin{align}\label{eq:action_fL}
\frac{S[f,\Lambda]}{N} &= K \int d\tau \left[\Lambda'(\tau)^2-(\mu f'(\tau))^2\right] - 2\alpha_{S} \int d\tau \left\lbrace \tan \left(\frac{f(\tau)}{2} \right),\tau \right\rbrace +\frac{\kappa}{2^{2\Delta-1}}\int d\tau\,\left( f'(\tau)\right)^{2\Delta} \,,
\end{align}
where
\begin{align}\label{eq:kappa_def}
\kappa=|\eta|\,,
\end{align}
and we have absorbed the phase $\theta$ of $\eta$ in the constant we introduced in \eqref{eq:f_lambda_ansatz}. 
We shall show numerically later that, for sufficiently low temperatures, corresponding to the wormhole phase $\mathcal{E} \approx \mu \beta $. The prefactor $K$ is the compressibility that may also depend on microscopic parameters such as the chemical potential.

Following the procedure of Ref.~\cite{maldacena2018} for the Majorana case, we can see that the solution
\begin{align}\label{eq:eternal_wormhole}
f(\tau) = \textbf{t}' \tau\,, \qquad \Lambda =0\,,
\end{align}
with
\begin{align}
\textbf{t}'^{2(1-\Delta)} = \frac{\eta\Delta}{2^{2\Delta} (\alpha_{S}-K\mu^2)}\,,
\end{align}
satisfies the equations of motion and leads to a vanishing Noether charge for the global time translation symmetry. Is the constraint from the Noether charge that fixes the value of $\textbf{t}$'. This is precisely the uncharged eternal wormhole of \cite{maldacena2018}.

In order to compute the physical Green's functions from \eqref{eq:c_IR_saddleLLRR}, \eqref{eq:c_IR_saddleLRRL} we perform a reparametrization and gauge transformation with the classical solution \eqref{eq:eternal_wormhole}, leading to
\begin{align}\label{eq:saddle_wormhole}
G_{LL}(\tau) =G_{RR}(\tau) \sim \left[\frac{\textbf{t}'}{ 2 \sin \frac{\textbf{t}' \tau}{2} } \right]^{2\Delta} \,,\qquad G_{LR}(\tau) \sim G_{RL}(\tau) \sim \left[\frac{\textbf{t}'}{ 2 \cos \frac{\textbf{t}' \tau}{2} } \right]^{2\Delta}\,,
\end{align}
The value of $\textbf t'$ determines the energy scale of the conformal excitations of the model. With respect to the boundary time, and assuming that this is the lowest excitation, the energy gap is $E_g \sim \textbf t'\Delta/\alpha_S$ and the spectrum of low energy excitations is linear \cite{maldacena2018}.  

We postpone a detailed study of the Noether charges, the quantization of (\ref{eq:action_fL}), the solution of the associated Liouville quantum mechanical problem and its derivation from a gravity dual to a future publication \cite{usfuture}. 
\subsection{Numerical evaluation of $\mathcal{E}$ and qualitative phase diagram}
In order to gain further insight about the low energy effective action (\ref{eq:action_fL}), 
we carry out the numerical evaluation of the parameter $\mathcal{E}$  by fitting the numerical Green's function obtained in section \ref{sec:fivenum} with the 
ansatz (\ref{eq:c_IR_saddleLLRR}), (\ref{eq:c_IR_saddleLRRL}). We restrict ourselves to the wormhole and intermediate phase as this ansatz will work only for sufficiently low temperatures. 

\begin{figure}
	\includegraphics[scale=.7]{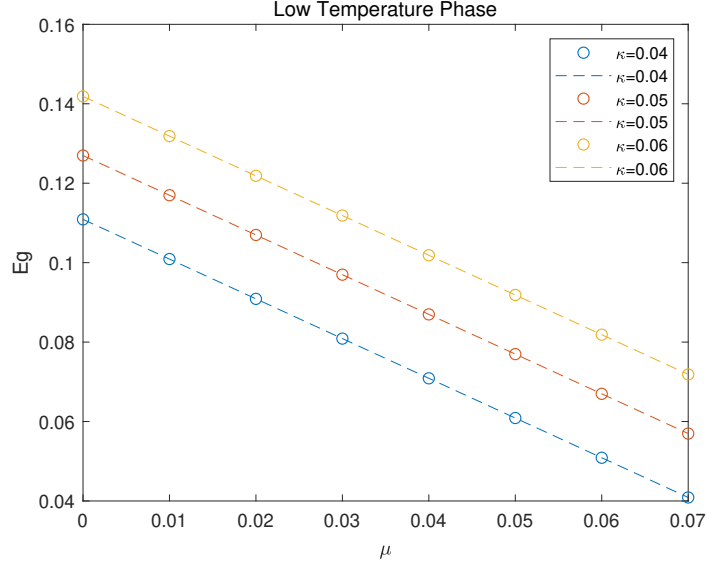}
	\caption{The gap $E_g$ in the low temperature limit as a function of $\mu$ for different $\kappa$'s. The fitting $E_g = E_0(\kappa) - \mu$ is in excellent agreement with the numerical results.}\label{fig:egmu}
\end{figure}

In the wormhole phase, we have found that, with great accuracy, 
$\mathcal{E}T \approx \mu$ which implies that $E_g$ defined in previous section to characterize the energy gap between the ground state and first excited state in the cold wormhole phase has a simple relation with $\mu$, $E_g = E_0 - \mu$ where $E_0 > \mu$ only depends on $\kappa$ and $J$ and therefore it is a more accurate indicator of the wormhole phase. In figure \ref{fig:egmu}, we depict result of the $\mu$ dependence of $E_g$ that confirm this simple relation. 

In the intermediate phase, a similar fitting of Green's function to the ansatz (\ref{eq:c_IR_saddleLLRR}), (\ref{eq:c_IR_saddleLRRL}) points to a relation $T \mathcal{E} \sim b + cQ\mu$ with $b\sim \mu$ and $c$ a numerical factor of order one. However, due to the relatively narrow window of parameters, our results are less reliable than those in the cold wormhole phase so this expression for $\mathcal{E}$ must be considered more like a conjecture that requires further verification.

In any case, it seems that, as in the canonical ensemble, there is a close relation between the $\mathcal{E}$ and the charge $Q$. Indirectly, this is another indication that the intermediate phase is still a wormhole phase as $E_0$ is not much perturbed by a finite charge, namely, the presence of a finite chemical potential and charge will reduce the gap induced by the coupling of left and right SYK's but provided that $E_0(\kappa)$ is finite and the total gap $E_g$ does not vanish, it seems that these are separate effects and that the physics of the wormhole is not qualitatively altered. 

This view is reinforced by a qualitative analysis of the low energy effective action. In the large $N$ limit and at finite physical temperature, the saddle point solution for $\Lambda$ is $\Lambda = 0$ 
. The low energy effective action has a contribution proportional to $f'^2(\tau)$. Since $T \mathcal{E} \approx \mu$, the effect of a small chemical potential in the Schwarzian action is a small renormalization of $\alpha_S \to \alpha_S -K\mu^2$ that will change slightly the specific heat. 
In the intermediate region, the charge jumps to a finite value and then increase linearly with temperature so this renormalization becomes increasingly important and eventually will destabilize the wormhole phase for sufficiently large $\mu$ or higher temperature. However, if the increase of $Q$ is sufficiently small at the transition, the gap is finite and the wormhole phase may survive. 

Another path to show the existence of the two transitions comes from a qualitative estimation of the critical temperatures from the effective low energy grand potential. 
At the critical temperature, the grand potential of the two phases must be the same. The black hole high temperature phase is approximately given by two times  (\ref{eq:grand_potential_thermal}), the low temperature wormhole phase is given by $E_g$. In the intermediate phase, a finite but smaller $E_g$ may survive, $Q$ jumps at the two transitions and becomes linear in temperature in between. For the intermediate phase to be a charged wormhole, the zero temperature entropy $\mathcal S(\mathcal{E})$ must remain zero. 
Therefore, for the intermediate phase to be some kind of charged wormhole, we would suggest that the critical temperature of the two phase transitions can be estimated by,
\begin{align}
|E_g| \sim 4\pi Q\mathcal{E}\,, \qquad 4\pi Q\mathcal{E} \sim 2\mathcal{S}(\mathcal{E})\,.
\end{align}

\section{Conclusions and outlook}\label{sec:conclusion}
We have studied a coupled two-site SYK model with Dirac fermions. Many of the features of this model are qualitatively similar to the analogous model with Majorana fermions. For sufficiently small chemical potential, 
the ground state is gapped with a value that decreases with the chemical potential. It is likely dual to an eternal traversable with zero charge wormhole despite of the presence of a finite chemical potential. As temperature increases, and for a small coupling between the two SYKs, eventually we observe a first order transition from the wormhole phase to likely the black hole phase. As the coupling increases, the first order transition eventually becomes a sharp crossover.

As the chemical potential increases, we have found there is an important qualitative difference with respect to the Majorana case: we have identified a range of weak couplings and not too small chemical potentials for which an intermediate phase, tentatively termed charged wormhole phase, occurs. There is still a gap in the spectrum though the charge, which was zero in the wormhole phase, becomes suddenly finite. It is separated from the black hole phase by a first order transition at higher temperature. At this second critical temperature, the charge undergoes an additional abrupt increase. These transitions become crossovers for sufficiently large chemical potential or strong coupling between the left and right complex SYKs. \\
The thermodynamic features of the model, obtained from the numerical solution of the SD equations, are in qualitative agreement with results obtained from a low energy effective model based on the approximate conformal symmetry of the ground state, close to a charged TFD state. This effective model is a generalized coupled Schwarzian action with extended $SL(2,R)\times U(1)$ symmetry that reflects the additional charge degree of freedom.\\
Finally, we enumerate a few natural extensions of this work. A detailed study of the gravity dual of this model could shed additional light on the nature of the intermediate phase. More specifically, it would be interesting to derive the low energy effective action and the associated Liouville quantum mechanical problem starting from the gravity dual or to extend the novel boundary conditions in AdS$_2$ \cite{godet2020}, dual to a single complex SYK, to our coupled complex SYK model. 
It would also be worthwhile to compute transport properties such as the conductivity in order to further characterize the field theory dual of the wormhole phase. For that, it would also be necessary to generalize the model to higher spatial dimensions. That could bring closer an experimental realization of the physics of the SYK model and its gravity dual.  Other venues for further research includes the extension of these results to supersymmetric SYK models, non-random SYK models and a detailed description of the real-time formation of the wormhole by coupling the model to a thermal reservoir \cite{maldacena2019}.

{\it Note:} Near the completion of this work, the paper \cite{sahoo2020} was posted in arxiv that investigate the same model though most of the calculation were focused on the case of no chemical potential. As far as we know, the intermediate phase was not identified. For related work on this model, see also the recent papers \cite{Nedel:2020bxa,Sorokhaibam:2020ilg}.

\acknowledgments  We acknowledge financial support from a Shanghai talent program and from the National Natural Science
Foundation of China (NSFC) (Grant number 11874259)

\bibliographystyle{apsrev4-1}
\bibliography{library2}

\end{document}